\documentclass{aa_nonum}
\usepackage{graphicx}
\usepackage{txfonts}
\usepackage{url}
\usepackage{hyperref}
\hypersetup{
    colorlinks=true,
    linkcolor=blue,
    urlcolor=blue,
    citecolor=blue
}
\usepackage[flushleft]{threeparttable}

\usepackage{orcidlink}

\usepackage{array}
\newcolumntype{C}[1]{>{\centering\let\newline\\\arraybackslash\hspace{0pt}}m{#1}}

\usepackage{xspace}

\def\aj{AJ}

\def\apj{ApJ}
\def\apjl{ApJ}
\def\apjs{ApJS}
\def\ao{Appl.~Opt.}
\def\apss{Ap\&SS}
\def\aap{A\&A}

\def\aaps{A\&AS}

\def\mnras{MNRAS}

\def\pasp{PASP}
\def\pasj{PASJ}

\def\nat{Nature}

\def\physrep{Phys.~Rep.}

\newcommand{\ionic}[2]{#1$\,${\scshape{#2}}\xspace}%    % ion, i.e., C II = \ionic{C}{ii}

\begin{document}

   \title{Disentangling the X-ray variability in the Lyman continuum emitter Haro 11}

%\subtitle{}

   \author{A.~Danehkar \inst{1}\orcidlink{0000-0003-4552-5997} \and 
   S.~Silich \inst{2}\orcidlink{0000-0002-3814-5294} \and 
   E.~C.~Herenz \inst{3}\orcidlink{0000-0002-8505-4678} \and 
   G.~\"{O}stlin \inst{4}\orcidlink{0000-0002-3005-1349}
          }

   \institute{Eureka Scientific, 2452 Delmer Street, Suite 100, Oakland, CA 94602-3017, USA \label{inst1} \and 
   Instituto Nacional de Astrof\'{i}sica, \'{O}ptica y Electr\'{o}nica (INAOE), AP 51, 72000, Puebla, M\'{e}xico \label{inst2} \and 
   Inter-University Centre for Astronomy and Astrophysics (IUCAA), Pune University Campus, Pune 411007, India \label{inst3} \and  
   Department of Astronomy, Oskar Klein Centre, Stockholm University, AlbaNova, SE-106 91, Stockholm, Sweden \label{inst4} 
   \\ \email{danehkar@eurekasci.com}
             }
             
\date{Received 29 January 2024 / Accepted 19 June 2024 }

% \abstract{}{}{}{}{} 
% 5 {} token are mandatory

% Background: background information (present tense)
% Aim: principal activity (past/present perfect tense)
% Method: methodology (past tense)
% Results: results (past tense)
% Conclusion: conclusions (present tense/tentative verbs/model auxiliaries)

  \abstract
    {Lyman break analogs in the local Universe serve as counterparts to Lyman break galaxies (LBGs) at high redshifts, which are widely regarded as major contributors to cosmic reionization in the early stages of the Universe.
    }
   {We studied \textit{XMM-Newton} and \textit{Chandra} observations of the nearby LBG analog Haro\,11, which contains two X-ray-bright sources, X1 and X2. Both sources exhibit Lyman continuum (LyC) leakage, particularly X2.
   }
   { 
We analyzed the X-ray variability using principal component analysis (PCA) and performed spectral modeling of the X1 and X2 observations made with the \textit{Chandra} ACIS-S instrument.
   }
   {
   The PCA component, which contributes to the X-ray variability, is apparently associated with variable emission features, likely from ionized superwinds. Our spectral analysis of the \textit{Chandra} data indicates that the fainter X-ray source, X2 (X-ray luminosity $L_{\rm X} \sim 4 \times 10^{40} $\,erg\,s$^{-1}$), the one with higher LyC leakage, has a much lower absorbing column ($N_{\rm H} \sim 1.2 \times 10^{21}$\,cm$^{-2}$) than the heavily absorbed luminous source X1 ($L_{\rm X} \sim 9 \times 10^{40} $\,erg\,s$^{-1}$ and $N_{\rm H} \sim  11.5 \times 10^{21}$\,cm$^{-2}$).
   }
   {We conclude that X2 is likely less covered by absorbing material, which may be a result of powerful superwinds clearing galactic channels and facilitating the escape of LyC radiation. Much deeper X-ray observations are required to validate the presence of potential superwinds and determine their implications for the LyC escape.
   }


\titlerunning{X-ray Variability in Haro\,11}

\authorrunning{Danehkar et al.}
   \maketitle
%
%-------------------------------------------------------------------
%\section{Introduction}
%
%
%
%
%
%\par ~~~
%
%\par ~~~

\section{Introduction}
\label{haro11:introduction}

% Move 1: General statements about the field of research to provide the reader with a setting for the problem to be report
% Move 2: More specific statements about the aspects of the problem already stuidied by other researchers
% Move 3: gap + reearch topic: statements that indicate the need for more investigation
% Move 4: Statement of purpose: research orientation:
% very specific statements giving the purpose/objective of the writer's study

X-ray observations of Lyman continuum (LyC) emitters in the early Universe face significant challenges due to their considerable distance. Our understanding of their X-ray characteristics is inferred from stacking analyses \citep{Nandra2002,Lehmer2005,Laird2006,Zinn2012,Basu-Zych2013a} and optimized averaging analysis \citep{Cowie2012}. The high redshifts of distant galaxies result in the rest-frame X-rays shifting into the energy range below a few keV. However, the extinction caused by the intergalactic medium, poor spatial resolutions, and low signal-to-noise ratios of distant galaxies preclude us from getting a complete picture of the X-ray properties of high-$z$ LyC emitters, which could contribute to the majority of the early Universe's ionizing budget \citep{Barkana2001,Robertson2010,Dayal2018}. Because of this complication, it is difficult to study the X-ray features of high-$z$ starburst galaxies. However, a detailed X-ray analysis of their nearby analogs helps us better understand the X-ray properties of LyC emitters in the faraway Universe. 

\begin{figure}%[htbp]
   \centering
%\medskip
\includegraphics[width=3.0in, trim = 0 0 0 0, clip, angle=0]{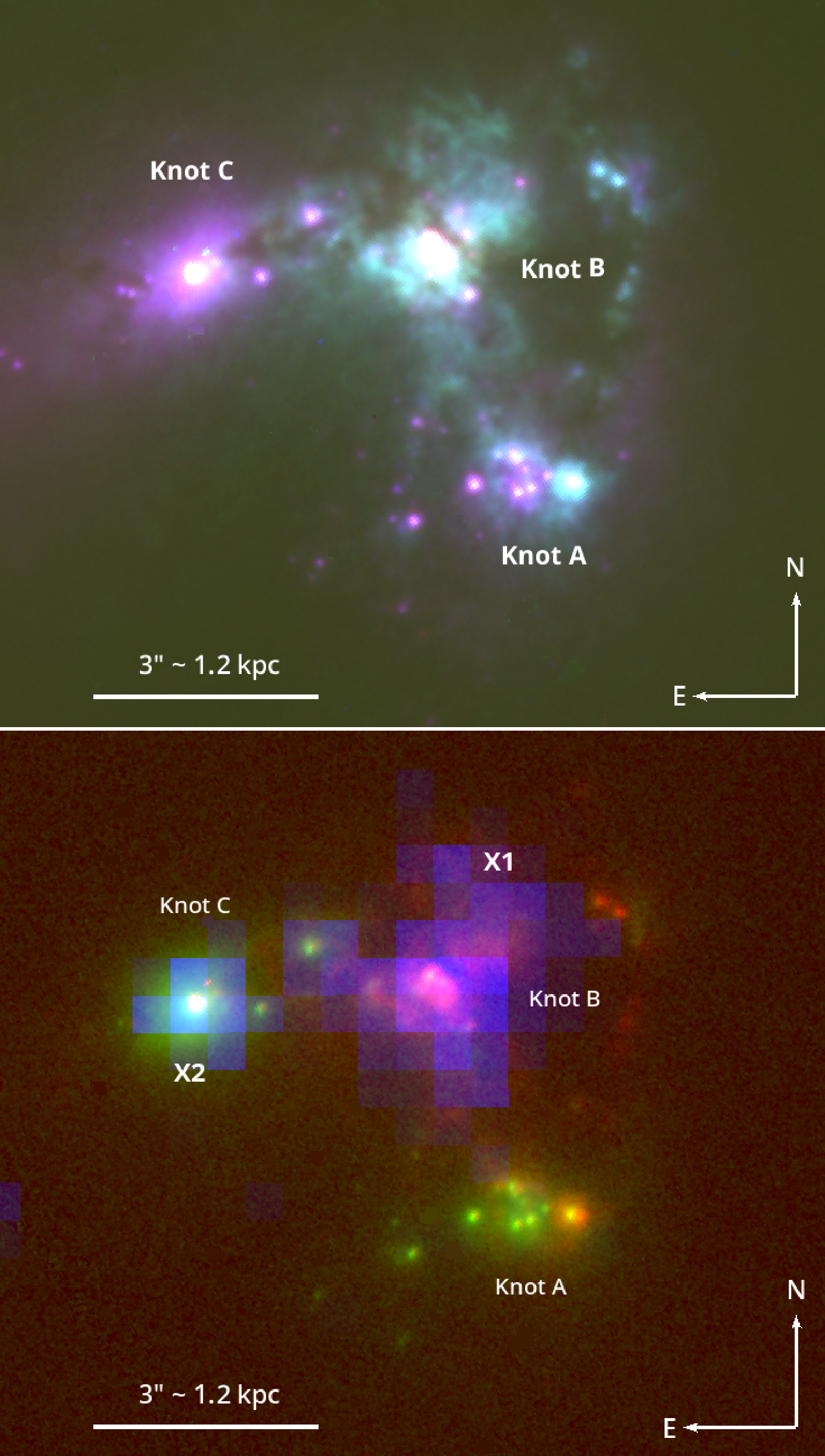}
%\medskip
%\vspace{-5pt}
\caption{Multiband composite images of Haro\,11. Top panel: HST WFC3/UVIS observations using the F555W (blue), F665N (green), and F814W (red) filters (Prop.\,ID 15649, PI. Chandar). \textit{Bottom panel:} Multiband imaging in H$\alpha$ emission (HST F665N, Prop.\,ID 15649; red), Ly$\alpha$ emission (HST ACS/SBC F122M, Prop.\,ID 9470; green), and the X-ray 3--5 keV band (\textit{Chandra} ACIS-S, Obs.\,ID 8175; blue). The HST data are available via MAST at doi:\href{https://doi.org/10.17909/rn22-jy61}{10.17909/rn22-jy61}.
\label{fig:haro11}
}
\end{figure}

Haro\,11 (ESO\,350-IG\,038) is the closest LyC leaking dwarf galaxy \citep{Bergvall2006,Leitet2011,Rivera-Thorsen2017,Oestlin2021,Komarova2024} in the local Universe \citep[$z = 0.0206$ and $D = 82$\,Mpc;][]{Micheva2010} that shares many characteristics with Lyman break galaxies (LBGs) in the early Universe \citep{Grimes2007}. It is also classified as a Ly$\alpha$-emitting source, and its young star-cluster population is similar to those of high-$z$ Ly$\alpha$ emitters \citep{Hayes2007}. The Ly$\alpha$ leakage in this galaxy has been confirmed by means of UV imaging \citep{Oestlin2009,Oestlin2021} and UV spectroscopy \citep{Rivera-Thorsen2017}. This compact blue galaxy has a high star formation rate of $\sim 20$--28.6 $M_{\odot}$\,yr$^{-1}$ \citep{Grimes2007,Madden2013} and a relatively low metallicity of $12+[$O/H$]=7.9$ \citep[][]{Oestlin2015}, similar to LBGs at high redshifts. It has been extensively studied in the IR, optical, UV, and X-ray \citep[e.g.,][]{Bergvall2000,Hayes2007,Oestlin2009,Adamo2010,Cormier2012,Prestwich2015,Gross2021}. It contains three optically bright predominant knots, labeled A, B, and C in the \textit{Hubble} Space Telescope (HST) observations (see Fig.~\ref{fig:haro11}, top). 
Each of these knots comprises several super star clusters (SSCs), with ages ranging from 1 to 40 Myr, and has an average star formation rate of around 22 $M_{\odot}$\,yr$^{-1}$ \citep{Adamo2010}. The LyC escape fraction was deduced to be around 30\% in Knot C, but much lower in Knot B \citep{Oestlin2021}. More recently, the LyC escape fractions within 903--912\,{\AA} were recently estimated to be $3.4 \pm 2.9$\% in Knot B and $5.1 \pm 4.3$\% in Knot C \citep{Komarova2024}. Haro\,11, which is analogous with high-$z$ starburst galaxies \citep{Sirressi2022}, allows us to study the X-ray characteristics of an LBG analog with LyC leakage that is similar to LBGs in the early Universe.

Knot\,B in Haro\,11 is coincident with the X-ray source X1 (CXOU\,J003652.4-333316; see Fig.~\ref{fig:haro11}), which contains extremely hard (3--5 keV) X-rays \citep[$L_{\rm X}\sim 10^{41}$\,erg\,s$^{-1}$;][]{Prestwich2015,Gross2021}. It is likely that the X-rays in X1 originate from an active galactic nucleus (AGN) or an extreme black hole binary  \citep[BHB;][]{Prestwich2015}. Knot\,B is surrounded by an extended superbubble visible in H$\alpha$ emission \citep{Menacho2019}, which is likely powered by the star cluster wind. X-ray observations also show that Knot\,C is coincident with the X-ray source X2 (CXOU\,J003652.7-333316;  Fig.~\ref{fig:haro11}, bottom), which could be associated with an ultra-luminous X-ray (ULX) source \citep[$L_{\rm X}\sim 5$--$6 \times 10^{40}$\,erg\,s$^{-1}$;][]{Gross2021}. 
More recently, \citet{Komarova2024} found that both X1 and X2 host ULXs. The X2 source could be capable of powering a strong superwind that could form channels in the surrounding interstellar medium (ISM) while creating a fragmented kiloparsec-scale superbubble \citep{Menacho2019}. 
In addition, the ISM around X2 was found to be less dense than around X1 \citep{Menacho2021} and to have a larger Ly$\alpha$ escape fraction \citep{Rivera-Thorsen2017,Oestlin2021} than the entire region \citep[$\sim$ 3 percent;][]{Hayes2007}. Moreover, the ISM around X2 was also found to be hotter ($\sim$ 16,600\,K) and with a lower metallicity (0.12\,Z$_{\odot}$) than that around X1 \citep[$\sim$ 12,500\,K, 0.35\,Z$_{\odot}$;][]{James2013}. Additionally, while X1 was seen to have a greater metal enrichment in integral field spectroscopic observations with the Multi Unit Spectroscopic Explorer (MUSE) on the Very Large Telescope (VLT), X2 was found to have higher temperature discrepancies, which could be partially due to shock ionization \citep{Menacho2021}.

Multiband imaging analysis of Haro\,11 by \citet{Adamo2010} revealed over 200 young and massive SSCs (masses in the range $10^{4}$--$10^{7}$\,M$_{\odot}$) that mostly emerged during the starburst that ignited about 40\,Myr ago. In addition, supernova explosions could have begun in about half of the SSCs around 3.5\,Myr ago. The frequent occurrence of supernova explosions, as found by \citet{Sirressi2022}, can lead to hot gas winds around the knots, which illuminate in the diffuse X-ray emission. Previously, \citet{Gross2021} classified the diffuse emission in the vicinity of Knots B and C as the ``AGN/composite'' region of the Baldwin--Phillips--Telervich (BPT) diagram. However, \citet{Sirressi2022} find that their emission lines are situated around the error margins of the line dividing starbursts from AGNs in the BPT diagram. As mentioned by \citet{Menacho2021}, the substantial energy deposition from stellar feedback and supernovae during the recent starburst can generate the observed outflows, resulting in the observed diffuse soft X-ray emission. \citet{Menacho2021} also conclude that Knot C (X2), home to about 80 percent of the supernovae that occurred in Haro 11 in the last 10\,Myrs, is the dominant source of energy released in supernova explosions in the system.

\begin{table*}
%\centering
\begin{center}
\caption[]{Observation log of Haro\,11.
\label{haro11:obs:log}}
\footnotesize
\begin{tabular}{lllllcrr}
\hline\hline\noalign{\smallskip}
Observatory     &Instrument     &Obs.\,Mode     &Filter/Op. Mode        &Obs.\,ID \& PI   &\multicolumn{1}{c}{Obs. Start (UTC)}   &Exp. (ks)\,$^{\rm \bf a}$      &Counts\,$^{\rm \bf a}$ \\
\noalign{\smallskip}
%\tableline
\hline
\noalign{\smallskip}
XMM     &EPIC-pn        &Imaging        &Thin-1 Filter  &0200800101, Wolter     &2004 Dec 14, 20:38   &31.29 [18.06]  & 1395 [588] \\
XMM     &EPIC-MOS2      &Imaging        &Thin-1 Filter  &0200800201, Wolter     &2005 May 21, 11:02   &59.91 [45.90]  & 884 [448] \\
CXO     &ACIS-S &Very Faint     &Timed Event    &8175, Grimes   &2006 Oct 28, 11:45       &54.01  & 970 \\
CXO     &ACIS-S &Faint  &Timed Event    &16695, Prestwich       &2015 Nov 29, 11:00       &24.74  & 388 \\
CXO     &ACIS-S &Faint  &Timed Event    &16696, Prestwich       &2016 Sept 12, 20:42       &24.74  & 292 \\
CXO     &ACIS-S &Faint  &Timed Event    &16697, Prestwich       &2017 Nov 24, 14:00       &23.76  & 307 \\
\noalign{\smallskip}\hline
\end{tabular}
\end{center}
\begin{tablenotes}
\footnotesize
\item[1]\textbf{Note.} $^{\rm \bf a}$ Source counts over 0.4--10\,keV for XMM and over 0.4--8\,keV for CXO.
The exposure times and the count values in the square brackets correspond to the flaring background-filtered XMM events.
\end{tablenotes}
\end{table*}

The X-ray variability in \textit{Chandra} observations of Haro\,11 has recently been investigated by \citet{Gross2021}, who argue for flaring and fading similar to what is seen in the young starburst region of M82 and possible links to the Ly$\alpha$ and LyC emission. As M82 ($D = 3.6$\,Mpc) is much closer than Haro\,11, it provides much better X-ray details. In particular, several H-like and He-like emission lines caused by X-ray outflows have been identified in X-ray observations of M82 \citep{Ranalli2008,Liu2011,Lopez2020}. Specifically, the link between X-ray variability and outflows has been found in some ULXs \citep{Pinto2017,Kosec2018} and quasars \citep{Parker2017,Parker2018}. Moreover, \citet{Heil2010} found time delays between the hard and soft energy bands in ULXs, which may be caused by different locations of stratification layers in the X-ray outflow leading to X-ray variability. To better evaluate the nature of the X-ray variability in Haro\,11 and the possible evidence for outflows, we used principal component analysis (PCA), which serves as a statistical instrument for investigating the variability possibly caused by distinctive ionization zones of potential X-ray outflows as suggested by \citet[][]{Parker2018}. In this study we conducted PCA, along with spectral analysis, to disentangle the spectral variability of the X-ray sources. In Sect. \ref{haro11:observation} we describe the data reduction.  In Sect. \ref{haro11:analysis} we analyze the light curves using the hardness ratios and statistical methods, and perform PCA on the X-ray variability. Additionally, we use absorbed spectral models to reproduce the continua and simulate emission lines in the spectra. Our findings are discussed in Sect. \ref{haro11:discussions}, which is followed by a conclusion in Sect. \ref{haro11:conclusion}.

%\vfill\break

\section{Data}
\label{haro11:observation}

Haro\,11 was observed with the \textit{XMM-Newton} telescope \citep{Jansen2001} in 2004 and 2005, followed by the \textit{Chandra} X-ray Observatory \citep[CXO;][]{Weisskopf2000,Weisskopf2002} in 2006 and 2015--17.  Table \ref{haro11:obs:log} presents the observation log, including the instruments, exposure times, observing and operating modes. We used observations from the European Photo Imaging Camera-pn \citep[EPIC-pn;][]{Strueder2001} and the Metal Oxide Semiconductor (MOS) camera \citep[][]{Turner2001} on XMM. The energy coverage includes the energies from 0.15 to 12\,keV with a point spread function (PSF) resolution of $6''$ (full width at half maximum).
We also retrieved observations made by the Advanced CCD Imaging Spectrometer S-array \citep[ACIS-S;][]{Garmire2003} from  
the \textit{Chandra} Data Archive (CDA). The \textit{Chandra} dataset is available via the CDA at doi:\href{https://doi.org/10.25574/cdc.219}{10.25574/cdc.219}. The ACIS-S, characterized by minimal instrumental background, is capable of covering an energy range from 0.2 to 10 keV with a spatial resolution of $0.492''$  (CCD pixel size).

We obtained the original data files (ODFs) of Haro\,11 from the \textit{XMM-Newton} Science Archive. The data were reprocessed with the XMM Science Analysis Software \citep[\textsc{sas} v\,20.0.0;][]{Gabriel2004} and the current calibration files (XMM-CCF-REL-391) available since 2022 October 25, and adhered to the guidelines outlined in the XMM ABC Guide.\footnote{\href{https://heasarc.gsfc.nasa.gov/docs/xmm/abc/}{https://heasarc.gsfc.nasa.gov/docs/xmm/abc/}} The calibration index files (CIF) and ODFs were produced with the \textsc{sas} tasks \textsf{cifbuild} and \textsf{odfingest}, correspondingly. The pn and MOS2 event files were generated via the \textsc{sas} tasks \textsf{epchain} and \textsf{emchain}, respectively. To clean flaring particle background, we excluded the largely contaminated events at the beginning ($\sim$\,3.5 hours) of each observation, as well as a few of the subsequent events with count rates higher than, respectively, 0.4 and 0.35 c\,s$^{-1}$ identified in the single-pixel (PATTERN\,$=$\,0) 100\,s-binned light curves of the 10--12\,keV band for pn  and $>10$\,keV for MOS2. Furthermore, the final cleaned events were subjected to filtration using the \textsc{sas} tool \textsf{evselect} in order to exclusively retain patterned events (pn: PATTERN\,$\leqslant$\,4 and MOS2: PATTERN\,$\leqslant$\,12) and the appropriate PI channels (pn: 200\,$<$\,PI\,$<$\,15000 and MOS2: 200\,$<$\,PI\,$<$\,12000), while excluding any defective pixels (FLAG\,$=$\,0). The \textsc{sas} tool \textsf{edetect\_chain} was used to locate X-ray sources in the data with the aid of the attitude file made with \textsf{atthkgen} and the band image created by \textsf{evselect}. The \textsc{sas} task \textsf{especget} was used to extract the source spectra, as well as the related background spectra, redistribution matrix files (RMF), and auxiliary response files (ARF). The source spectrum of each observation was obtained by extracting data from a circular aperture with a radius of $36''$ centered on the brightest source. In contrast, the background spectrum was collected from a circular region of the same size and located on the same chip, but positioned distant from the PSF source wings and devoid of any sources. Haro\,11 was located near to the edge of the detector only in the EPIC-pn data in 2004 and the MOS2 in 2005.

The \textit{Chandra} Interactive Analysis of Observations package \citep[\textsc{ciao} v\,4.15;][]{Fruscione2006} was used to acquire the ACIS-S observations of Haro\,11 from the CDA, and reprocess them using the CALDB files (v\,4.10.2). 
The tool \textsf{specextract} provided by \textsc{ciao} for imaging-mode and zeroth-order grating data was employed to create pulse height amplitude (PHA) files with \textsf{dmextract}, as well as their redistribution and response files using \textsf{mkrmf} and \textsf{mkarf}. We customized this tool (\textsf{correctpsf}=yes, \textsf{weight}=no) in order to apply an energy-dependent point-like PSF correction to the final ARF via the \textsc{ciao} task \textsf{arfcorr}. The spectra of the X-ray sources X1 and X2 labeled in Fig.~\ref{fig:haro11} were extracted from circular regions with radii of, respectively, $1.6''$ and $1.1''$ located on their emission peaks. The background was selected from a source-free circular region with a radius of $12''$. 

\section{Analysis}
\label{haro11:analysis}

\subsection{Timing analysis}
\label{haro11:timing}

\begin{figure*}
\begin{center}
\includegraphics[height=12.3cm, trim = 0 0 0 0, clip, angle=0]{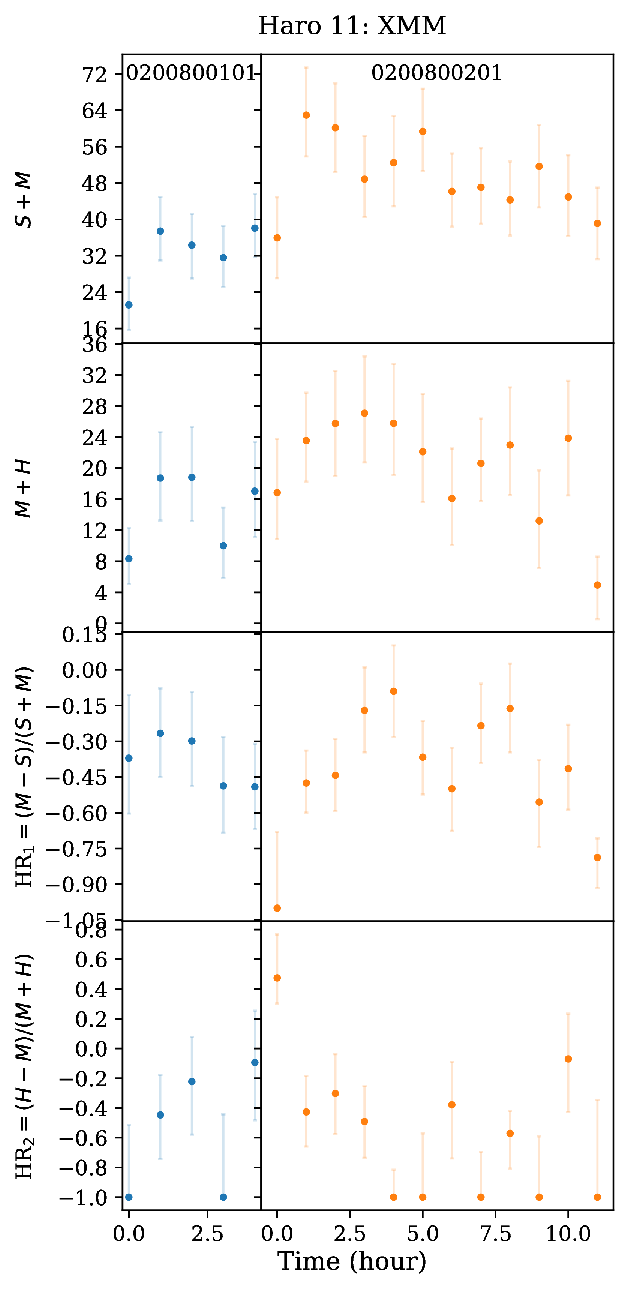}%
\includegraphics[height=12.3cm, trim = 0 0 0 0, clip, angle=0]{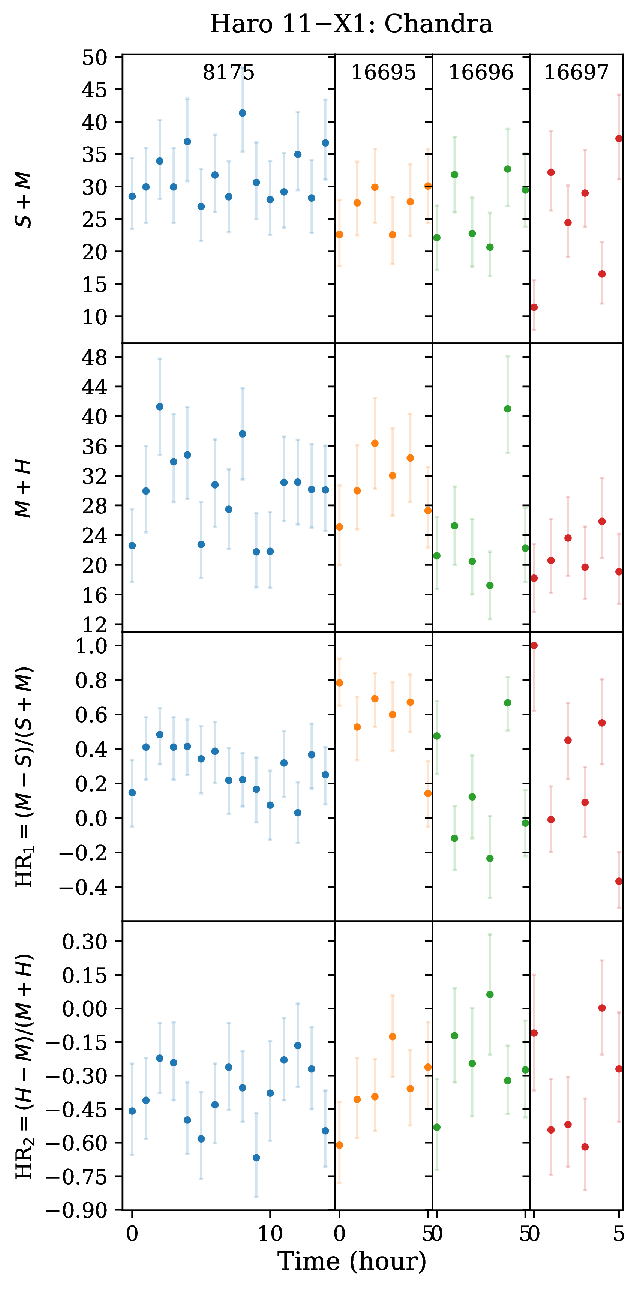}%
\includegraphics[height=12.3cm, trim = 0 0 0 0, clip, angle=0]{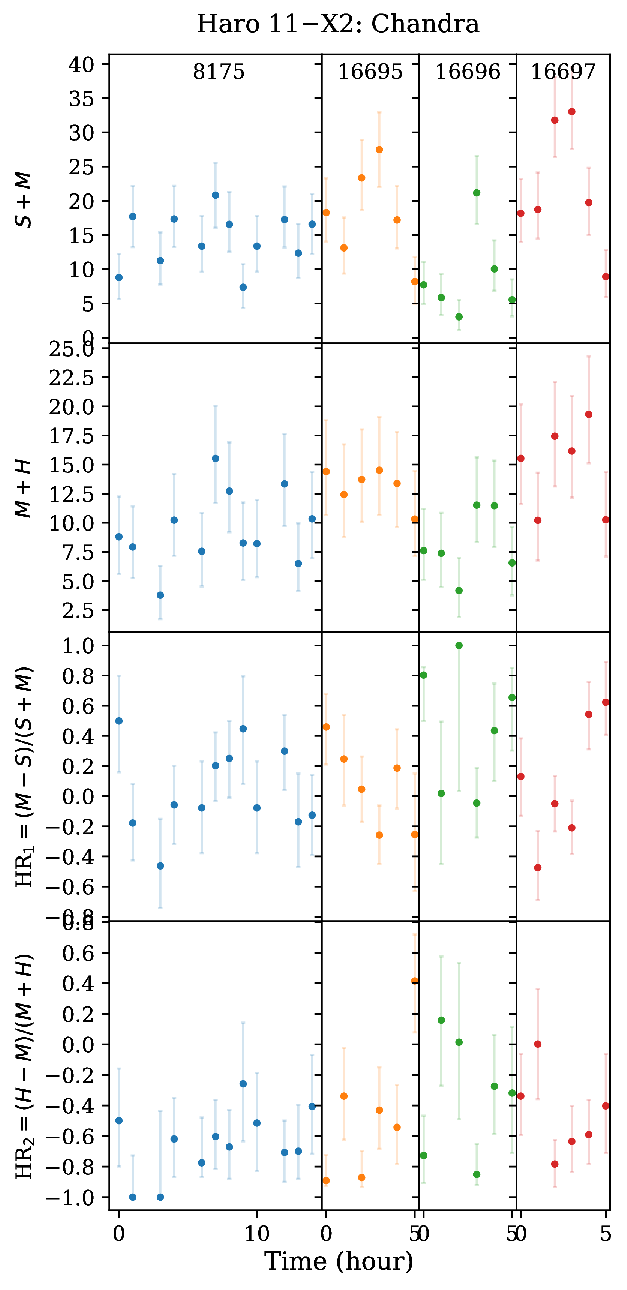}%
\end{center}
\caption{Background-subtracted light curves of Haro\,11 binned at 3.6\,ks in the $S+M$ and $M+H$ bands (in counts; upper panels), along with the hardness ratios ${\rm HR}_{1}$ and ${\rm HR}_{2}$ (lower panels) defined by Eqs. (\ref{eq_1}) obtained from the \textit{XMM-Newton} and \textit{Chandra} source and background light curves using the Bayesian quadrature method of the BEHR. The light curves of the later observations were scaled up based on the instrument effective areas relative to the first one.
The observation identifiers are given in the top panels.
\label{haro11:fig:lc}
}
\end{figure*}

To better evaluate the possible variability of the X-ray sources in this galaxy, we generated light curves over several energy ranges. 
We selected the soft ($S$: 0.4--1.1\,keV), the medium ($M$: 1.1--2.6\,keV), and the hard ($H$: 2.6--10\,keV) bands, with the exception of the \textit{Chandra} data, where the hard band was limited to 2.6--8\,keV.
The data series was discretized into time intervals of 1\,hour to ensure enough counts ($\gtrsim 5$) per bin, in addition to the clear capture of hourly variations. For the XMM data, the \textsc{sas} procedure \textsf{evselect} was used to extract light curves of the sources and backgrounds for the specified energy ranges. To compensate for the difference in the effective areas, the MOS2 light curves were scaled up using the ratio of the integrated effective areas of pn to MOS2 for each energy band.\footnote{$S_{\rm band}=\int_{E=E_{\rm min}}^{E_{\rm max}} A_{\rm eff} (E) dE$, where $A_{\rm eff}(E)$ is the effective area column from the ARF, and $E_{\rm min}$ and $E_{\rm max}$ are the lower and upper boundaries of the energy band, respectively.} Time-binned light curves of the sources resolved by the \textit{Chandra} telescope were produced with the \textsc{ciao} tool \textsf{dmextract} for X1 and X2 using the extraction apertures defined in Sec.\,\ref{haro11:observation}. To compensate for the decrease in the ACIS-S effectiveness over time, the light curves of the observations following the first one were scaled up according to the integration of the effective areas from the ARF over the specified energy range.
The soft excess of either an AGN or BHB, which may be present in X1, is usually characterized by a blackbody spectrum likely from the accretion disk. Moreover, powerful ionized superwinds can produce multicomponent warm absorbers, which  might be present in the medium band. The coronal, hot plasma within the innermost accretion flow around a black hole can also generate the continuum in the hard excess, commonly described by power law. 

To observe any transient changes in the spectral hardness, we also computed the hardness ratios $\mathrm{HR}_{1}$ and $\mathrm{HR}_{2}$ using the light curves of the soft ($S$), medium ($M$), and hard ($H$) bands ($H$) as follows:
\begin{align}
\mathrm{HR}_{1} = \frac{M-S}{S+M}, ~~~~ & \mathrm{HR}_{2} = \frac{H-M}{M+H}. \label{eq_1}
\end{align}
The similar approach was first introduced by \citet{Prestwich2003} for classifying X-ray sources. Hardness ratio analyses appear to be useful to constrain the actual nature of transient phenomena in various objects, such as low-mass X-ray binaries \citep[e.g.,][]{Finoguenov2002}, symbiotic binaries \citep{Danehkar2021a}, and characterize X-ray sources in starbursts \citep{Soria2003} and quasi-stellar objects \citep{Danehkar2018,Boissay-Malaquin2019}.

\begin{figure*}
\begin{center}
\includegraphics[height=11.3cm, trim = 0 0 0 0, clip, angle=0]{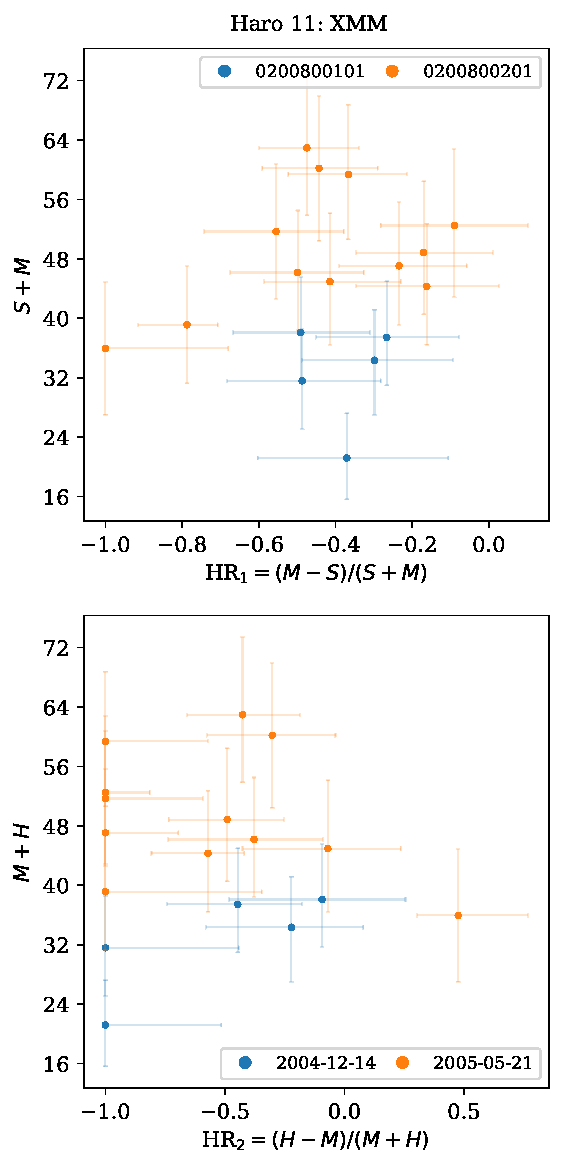}%
\includegraphics[height=11.3cm, trim = 0 0 0 0, clip, angle=0]{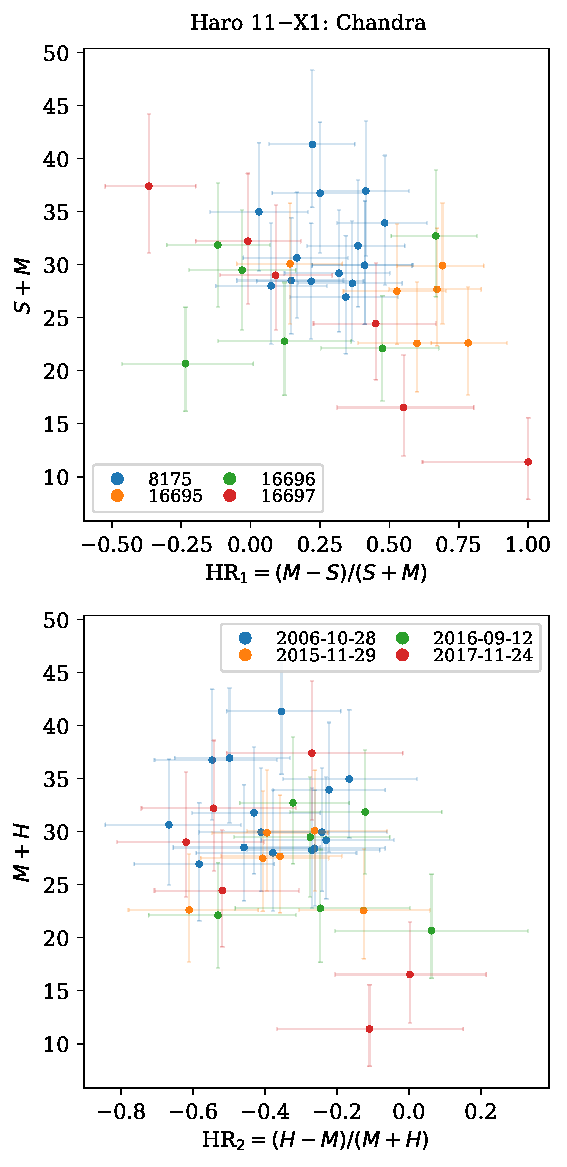}%
\includegraphics[height=11.3cm, trim = 0 0 0 0, clip, angle=0]{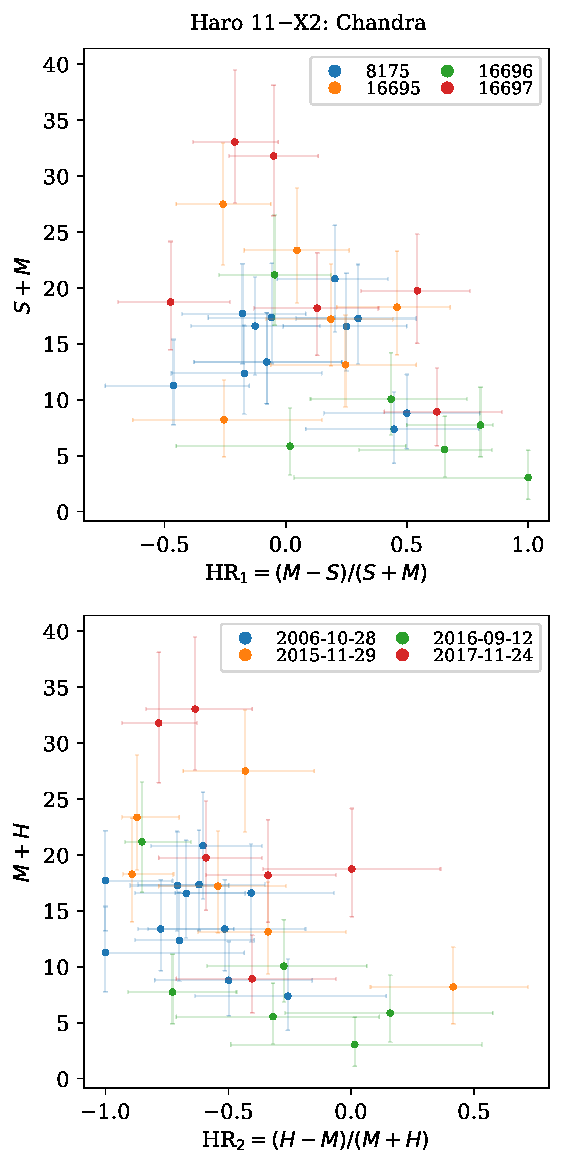}%
\end{center}
\caption{Hardness ratio diagrams of Haro\,11 made with \textit{XMM-Newton} and \textit{Chandra} observations. The hardness ratios ${\rm HR}_{1}$ (top) and ${\rm HR}_{2}$ (bottom) as defined by Eqs. (\ref{eq_1}) are plotted against the background-subtracted $S+M$ and $M+H$ bands (in counts), respectively, with a bin size of 3.6\,ks. 
The hardness ratios and the corresponding 1-$\sigma$ uncertainties were determined from the light curves of the source and background using the Bayesian quadrature method of the BEHR. The counts of the later observations were scaled up according to the instrument effective areas with respect to the first one.
\label{haro11:fig:hdr}
}
\end{figure*}

Figure~\ref{haro11:fig:lc} shows the light curves, % of Haro\,11 extracted from the \textit{XMM-Newton} data. 
including counts summed over $S +M$ and $M+ H$ bands (upper panels), and the hardness rations $\mathrm{HR}_{1}$  and $\mathrm{HR}_{2}$ (lower panels). 
The background-subtracted counts and the associated 1-$\sigma$ uncertainties were determined from the source and background light curves using 
the Gaussian quadrature method of the Bayesian Estimator for Hardness Ratios (BEHR) program \citep[][]{Park2006}. 
We see that Haro\,11 became slightly harder (higher $\mathrm{HR}_{2}$) occasionally during the XMM observations, albeit with high uncertainties. 
The brightness also started to increase (higher $S +M$ and $M +H$) and then gradually decrease in 2005. The \textit{Chandra} light curves show that the X1 source has higher counts than X2 in 2006 and 2016, though they have comparable counts later in 2017. While the hardness ratio $\mathrm{HR}_{1}$ is roughly similar in both X1 and X2, the $\mathrm{HR}_{2}$ is overall higher in X1 in 2006 and 2015 (i.e., a higher hardness). However, the source X2 looked occasionally harder than X1 during the 2016 observations. Some variability seen in X1 could be explained by the fact that it may host a BHB as suggested by \citet{Prestwich2015}. However, some temporary increases in brightness ($S +M$ and $M +H$) are also visible in X2, especially in 2015 and 2017.

Figure~\ref{haro11:fig:hdr} presents hardness ratio diagrams in which the hardness ratios $\mathrm{HR}_{1}$ and $\mathrm{HR}_{2}$ are plotted versus the light curves of the $S+M$ and $M+H$ bands %for the \textit{XMM-Newton} data 
binned at 3.6\,ks, respectively. The hardness diagrams from the 2005 XMM observations show some patterns, but they are not strong enough to be associated with any spectral-state transition similar to those seen in BHB \citep[see, e.g.,][]{Homan2005}. However, we find statistically significant changes in the HR$_2$ hardness ratio in the next subsection, which could be due to potential superwinds. 
Figure~\ref{haro11:fig:lc} shows that $\mathrm{HR}_{1}$ sometimes decreases with an increase in $\mathrm{HR}_{2}$. Moreover, the hardness ratio $\mathrm{HR}_{2}$ was occasionally lower when the source was softer and brighter (higher $S +M$ and $M +H$). The $\mathrm{HR}_{2}$ diagram of X1 suggests the possibility of some transition patterns between the ``faint/hard'' and ``bright/soft'' spectral states. This might be related to a possible BHB, though the diagrams lack vigorous transition patterns. 
The hardness diagrams of X2 depict robust transitions, which are results of occasional increases in brightness (higher $S +M$ and $M +H$) as seen in Fig.~\ref{haro11:fig:lc}.
Although the sources X1 and X2 cannot be separated in the \textit{XMM-Newton} observations, the brightening incidents seen in the \textit{Chandra} light curves suggest that both X1 and X2 could be responsible for the transient shapes created in the \textit{XMM-Newton} hardness diagrams, while the X1 source emits more X-ray photons than X2. 

%\vfill\null
%\vfill\eject
%\vfill\break

%\vfill\null
%\vfill\eject
%\vfill\break

\subsection{Statistical analysis of variability}
\label{haro11:statistic}

To quantify variability, we analyzed the light curves and their corresponding hardness ratios using different statistical methods for homogeneity and normality. 

To evaluate the homogeneity of time series, we calculated the von Neumann ratio defined as the mean squared successive difference with respect to the variance, $\eta \equiv \delta^2/\sigma^2$ \citep{vonNeumann1941},
\begin{align}
\delta^2 = \sum_{n=1}^{N-1}(x_{n+1}-x_n)^2/(N-1), ~~~~ & \sigma^2 = \sum_{n=1}^{N}(x_n-\mu)^2/(N-1), \label{eq_3}
\end{align}
where $\delta^2$ and $\sigma^2$ are the mean squared successive difference and the variance of data points, respectively, $n$ is the number of each data point, and $N$ is the total number of data points. The mean von Neumann ratio is $\bar{\eta}_{\rm norm}=2N/(N-1) \sim 2$ for a normal distribution \citep{Young1941}. We estimated the confidence levels on $\bar{\eta}_{\rm norm}$ according to the statistically significant threshold of $\alpha=0.05$. The von Neumann ratio of time series within the confidence intervals of the normal von Neumann ratio implies homogeneity, whereas those outside the aforementioned intervals suggest inhomogeneity in successive time series.

\begin{table*}
%\centering
\begin{center}
\caption[]{Statistical tests of X-ray variability.
\label{rtcru:stat:result}}
%\footnotesize
\begin{tabular}{lccccccccc}
\hline\hline\noalign{\smallskip}
Param.     &$\eta$      &$\bar{\eta}_{\rm norm}$    &\multicolumn{2}{c}{Lilliefors test}   &\multicolumn{2}{c}{AD test}   &\multicolumn{2}{c}{SW test}   & GL algo.  \\
     &      &    &$D$      &$p$-value   &$A^2$     &$p$-value   &$W $     &$p$-value   & $p_{\rm per}(\omega)$  \\
\noalign{\smallskip}
%\tableline
\hline
\noalign{\smallskip}
\multicolumn{10}{c}{XMM pn+M2} \\
\noalign{\smallskip}
S+M         & $    0.852 $ & $    2.125 \pm   0.951 $ & $    0.097 $ & $    0.941 $ & $    0.174 $ & $    0.912 $ & $    0.978 $ & $    0.936 $ & $    0.884 $ \\
M+H         & $    1.479 $ & $    2.125 \pm   0.951 $ & $    0.120 $ & $    0.732 $ & $    0.363 $ & $    0.400 $ & $    0.940 $ & $    0.314 $ & $    0.908 $ \\
HR$_1$        & $    1.477 $ & $    2.125 \pm   0.951 $ & $    0.185 $ & $    0.131 $ & $    0.468 $ & $    0.218 $ & $    0.923 $ & $    0.168 $ & \ldots \\
HR$_2$        & $    2.069 $ & $    2.125 \pm   0.951 $ & $    0.251 $ & $    0.006 $ & $    0.894 $ & $    0.017 $ & $    0.862 $ & $    0.016 $ & \ldots \\
\noalign{\smallskip}
\multicolumn{10}{c}{CXO X1} \\
\noalign{\smallskip}
S+M         & $    2.444 $ & $    2.062 \pm   0.682 $ & $    0.150 $ & $    0.061 $ & $    0.552 $ & $    0.143 $ & $    0.961 $ & $    0.275 $ & $    0.867 $ \\
M+H         & $    1.613 $ & $    2.062 \pm   0.682 $ & $    0.126 $ & $    0.214 $ & $    0.471 $ & $    0.230 $ & $    0.955 $ & $    0.187 $ & $    0.616 $ \\
HR$_1$        & $    2.325 $ & $    2.062 \pm   0.682 $ & $    0.059 $ & $    0.990 $ & $    0.111 $ & $    0.992 $ & $    0.995 $ & $    1.000 $ & \ldots \\
HR$_2$        & $    1.788 $ & $    2.062 \pm   0.682 $ & $    0.105 $ & $    0.470 $ & $    0.241 $ & $    0.754 $ & $    0.978 $ & $    0.730 $ & \ldots \\
\noalign{\smallskip}
\multicolumn{10}{c}{CXO X2} \\
\noalign{\smallskip}
S+M         & $    1.221 $ & $    2.069 \pm   0.716 $ & $    0.096 $ & $    0.669 $ & $    0.419 $ & $    0.308 $ & $    0.956 $ & $    0.246 $ & $    0.712 $ \\
M+H         & $    1.502 $ & $    2.069 \pm   0.716 $ & $    0.100 $ & $    0.613 $ & $    0.220 $ & $    0.820 $ & $    0.982 $ & $    0.882 $ & $    0.931 $ \\
HR$_1$        & $    2.045 $ & $    2.069 \pm   0.716 $ & $    0.132 $ & $    0.207 $ & $    0.345 $ & $    0.462 $ & $    0.970 $ & $    0.548 $ & \ldots \\
HR$_2$        & $    2.122 $ & $    2.069 \pm   0.716 $ & $    0.100 $ & $    0.606 $ & $    0.485 $ & $    0.210 $ & $    0.944 $ & $    0.119 $ & \ldots \\
\noalign{\smallskip}\hline
\end{tabular}
\end{center}
\begin{tablenotes}
\footnotesize
\item[1]\textbf{Notes.} $\eta$ is the von Neumann ratio, and $\bar{\eta}_{\rm norm}$ is the mean von Neumann ratio for homogeneous series. Normality tests include the Lilliefors, Anderson--Darling (AD), and Shapiro--Wilk (SW) methods, whose $p$-values examine the null hypothesis of a normal distribution. The Gregory--Loredo (GL) algorithm yields the probability $p_{\rm per} (\omega)$ of a periodic signal.
\end{tablenotes}
\end{table*}

To examine the normality of data points, we conducted three nonparametric statistical analyses of light curves, namely the Lilliefors test, the Anderson--Darling (AD) test, and the Shapiro--Wilk (SW) test. The Lilliefors test, which is an extension of the Kolmogorov--Smirnov (KS) approach, uses estimated mean $\mu$ and variance $\sigma^2$ of the sample for the examination of a normal distribution $\mathcal{N}(\mu,\sigma^2)$. The KS test is suitable for the standard normal distribution $\mathcal{N}(0,1)$. The Lilliefors test is based on the maximum difference between the empirical distribution function (EDF) of the sample and the cumulative distribution function (CDF) of the normal distribution \citep{Lilliefors1967}: 
\begin{align}
D = {\rm max}_{x}|F_{\rm norm}(x)-F(x)|, \label{eq_5}
\end{align}
where $F_{\rm norm}(x)$ is the CDF of the normal distribution $\mathcal{N}(\mu,\sigma^2)$, and $F(x)$ is the EDF of the sample. This test was conducted with the corresponding function from the \textsf{Statsmodels} package \citep{Seabold2010}. The AD test, which was also carried out with the relevant function from \textsf{Statsmodels}, is based on the weighted squared difference ($A^2$) between the EDF and the CDF, giving more weight to the tails of the CDF \citep{Anderson1952}:
\begin{align}
A^2 = N \int_{-\infty}^{\infty} [F(x)-F_{\rm norm}(x)]^2 w_{\rm norm}(x) dF_{\rm norm}(x), \label{eq_6}
\end{align}
where $w_{\rm norm}(x)\equiv[F_{\rm norm}(x) (1-F_{\rm norm}(x))]^{-1}$ is the weighting function. The SW test, which was conducted with the corresponding function from the \textsf{SciPy} package \citep{Virtanen2020}, utilizes the order statistics \citep{Shapiro1965}: 
\begin{align}
W = \left. {\left(\sum_{n=1}^{N} a_n x_n\right)^2}\middle/{\sum_{n=1}^{N} \left(x_n - \mu\right)^2} \right., \label{eq_7}
\end{align}
where $x_n$ are the sorted data points ($x_1<\cdots<x_n$), $\mu$ is the mean of data points, and $\mathbf{a}=(a_1,\ldots,a_n)$ is the coefficient vector defined as $\mathbf{a} \equiv \mathbf{m}^{\rm T} \mathbf{V}^{-1} / ( \mathbf{m}^{\rm T} \mathbf{V}^{-1} \mathbf{V}^{-1} \mathbf{m})^{\frac{1}{2}}$ using the vector $\mathbf{m}=(m_1,\ldots,m_n)^{\rm T}$ of the ordered values $m_n$ of the normal statistics and their covariance matrix $\mathbf{V}$. For all three statistical methods, $p$-value $\leqslant 0.05$ rejects the hypothesis of a normal distribution with significant statistics, whereas $p$-values between 0.05 and 0.1 are marginally significant for the null hypothesis. The SW was found to be the most effective method, followed by the AD and then Lilliefors \citep{Stephens1974,Razali2011}, though the AD is more powerful in the CDF with a sharp-edged peak and abruptly ending tails.

In addition, we employed the Gregory-Loredo (GL) algorithm to calculate the probability of a periodic signal, $p_{\rm per}(\omega)=O_{\rm per}(\omega)/[1+O_{\rm per}(\omega)]$, according to the sum of odds ratios $O_{\rm per}(\omega)=\sum_{m=2}^{m_{\rm max}} O_{m1}(\omega)$, where the odds ratio $O_{m1}(\omega)$ is defined by Eq. 5.25 in \citet{Gregory1992}. For $p_{\rm per} <  0.5$, there is definitely no variable, while $p_{\rm per}\geqslant 0.9$ is certainly variable. A light curve may be variable if $0.5\leqslant p_{\rm per} < 2/3$, and is considered to be variable if $2/3 \leqslant p_{\rm per}< 0.9$.

\begin{figure}
\begin{center}
\includegraphics[height=6.0cm, trim = 0 0 0 0, clip, angle=0]{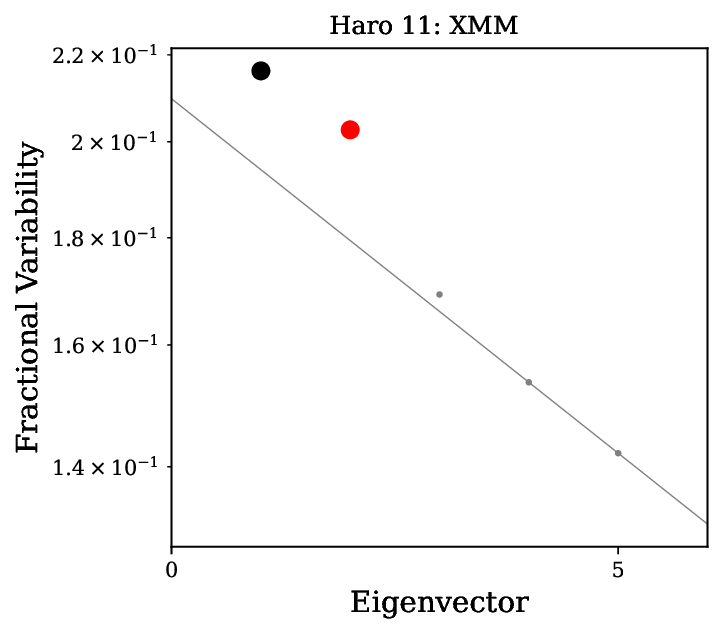}\\
\includegraphics[height=6.0cm, trim = 0 0 0 0, clip, angle=0]{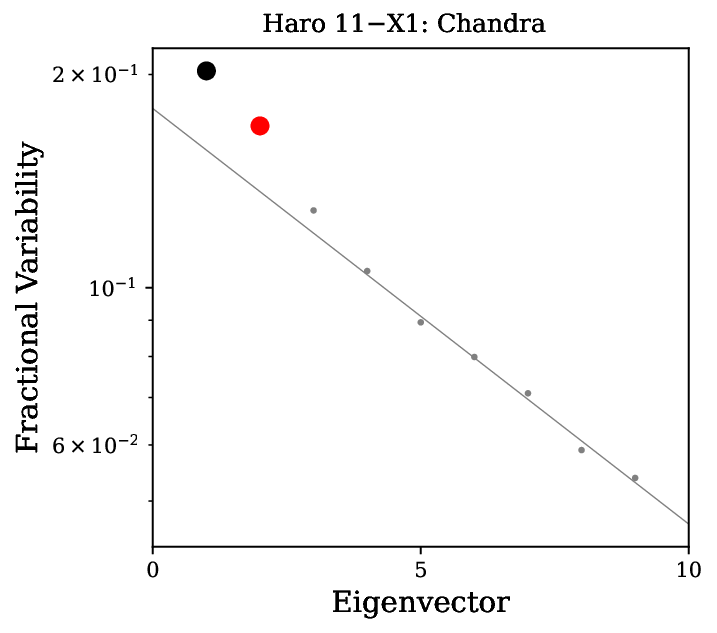}\\
\includegraphics[height=6.0cm, trim = 0 0 0 0, clip, angle=0]{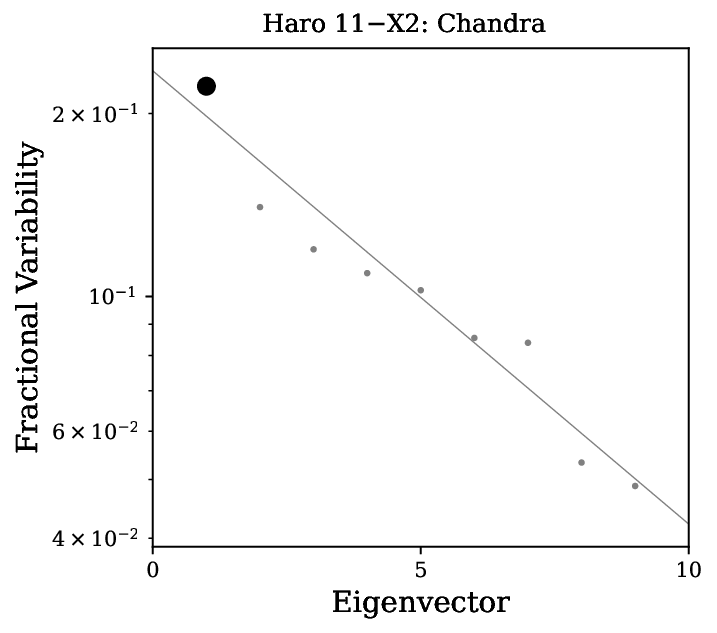}
\end{center}
\caption{Eigenvalue--fraction diagram of the \textit{XMM-Newton} EPIC-pn+MOS2 (top), \textit{Chandra} ACIS-S X1 (middle), and X2 (bottom). The normalized fraction variability values of PCA components are plotted against the corresponding eigenvector orders. The linear regression among logarithmic fractional eigenvalues starting from three onward is also plotted (gray line). The number of nonzero eigenvalues in each diagram corresponds to the number of time-sliced spectra.
The spectra and time series of the PCA components (color circles), which are statistically significant, are plotted with the same colors in Figure~\ref{haro11:fig:pca:1}.
\label{haro11:fig:pca:3}
}
\end{figure}

Table~\ref{rtcru:stat:result} presents our results derived from the aforementioned statistical methods. It can be seen that the von Neumann ratios ($\eta$) suggest inhomogeneity only in successive data points of the $S+M$ band for the XMM and CXO-X2 data, having $\eta$ outside the confidence intervals of $\bar{\eta}_{\rm norm}\sim 2$. We also notice the absence of normality with a significant statistic in HR$_2$ of the XMM data based on $p$-value $\leqslant 0.05$ from all three methods. In addition, there could be no normality in the $S+M$ light curve of the CXO-X1 observations with a marginally significant statistic according to $p$-value $\leqslant 0.1$ from the Lilliefors test. The $p$-values derived for the rest of the data are consistent with normality. The GL probability results of the data indicate definitely variability in $M+H$ of the XMM and CXO-X2 data. The $S+M$ data of XMM, as well as CXO X1 and X2, are also considered to be variable according to the values of $p_{\rm per}(\omega)$. The $M+H$ band in CXO X1 may be variable according to the GL probability outcome. However, we caution that the GL results could be less reliable due to the low number of data points. 

In summary, the spectral transition pattern seen in HR$_2$ of the XMM data in Figure~\ref{haro11:fig:hdr} is statistically significant according to normality tests, whereas hardness transitions are insignificant in other cases, apart from the marginal pattern in the $S+M$ band in X1. Because the XMM observations cover both X1 and X2, we cannot determine the origin of such a spectral transition in HR$_2$. Our analysis also yields $p$-values of around 0.2 and 0.1 with the AD and SW tests on HR$_2$ of the CXO X2 data, respectively, as well as $p$-values around 0.1--0.2 using the Lilliefors test on HR$_1$ of X2 and with all tests on HR$_1$ of XMM. The $p$-values are all small, thus perhaps suggestive of possible hardness transitions. Moreover, the $S+M$ light curves in XMM and CXO X2 do not appear to be homogeneous based on the von Neumann statistics, which again suggests the likelihood of nonrandom variations.

\subsection{Principal component analysis}
\label{haro11:pca}

\begin{figure*}
\begin{center}
\includegraphics[width=0.75\textwidth, trim = 0 0 0 0, clip, angle=0]{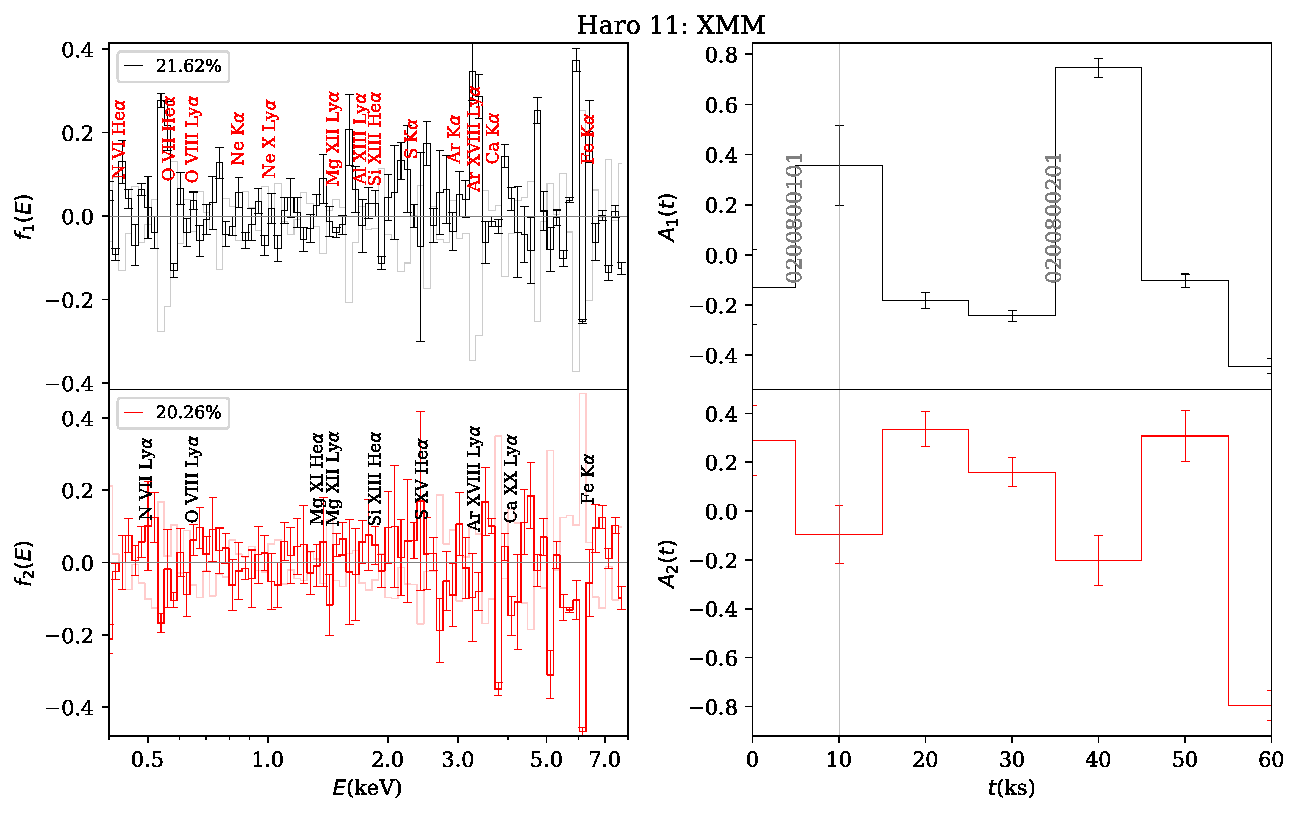}\\
\includegraphics[width=0.75\textwidth, trim = 0 0 0 0, clip, angle=0]{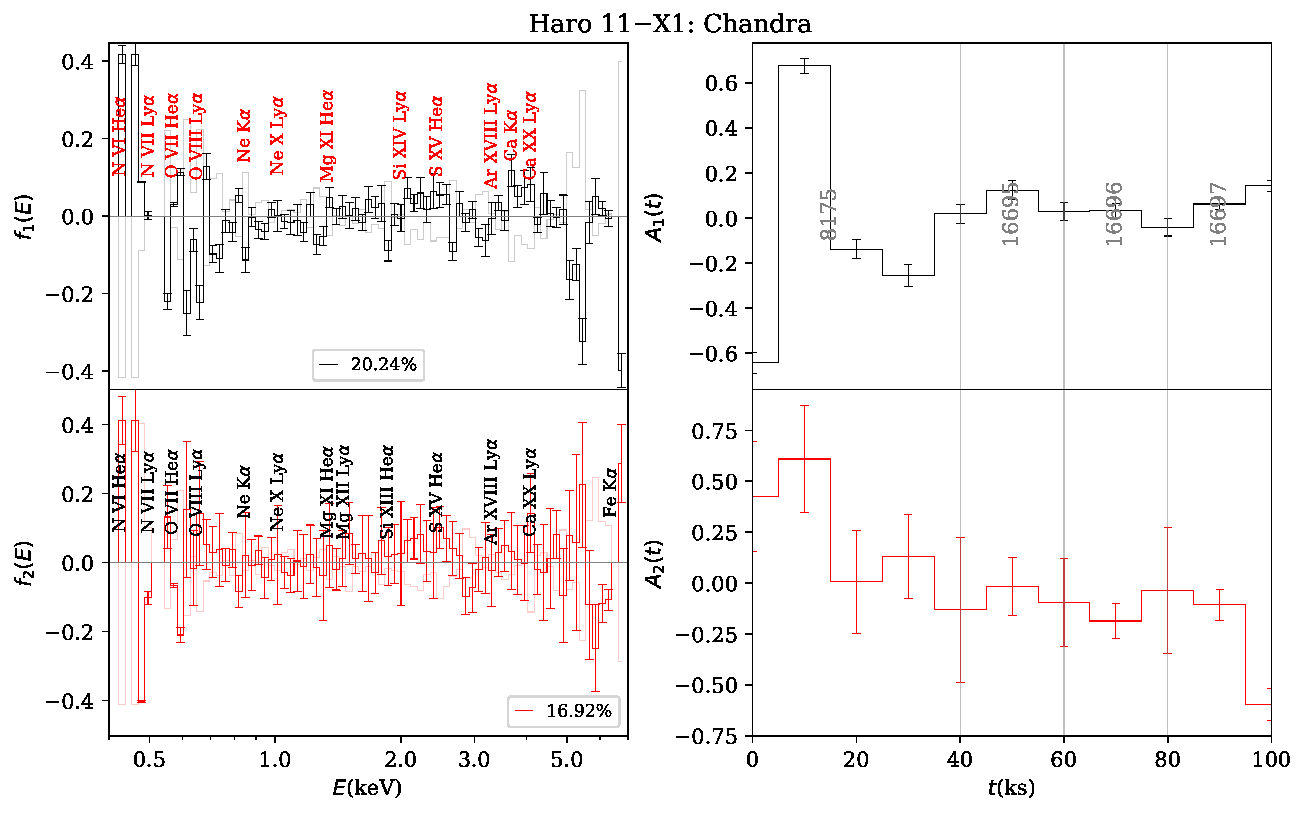}\\
\includegraphics[width=0.75\textwidth, trim = 0 0 0 0, clip, angle=0]{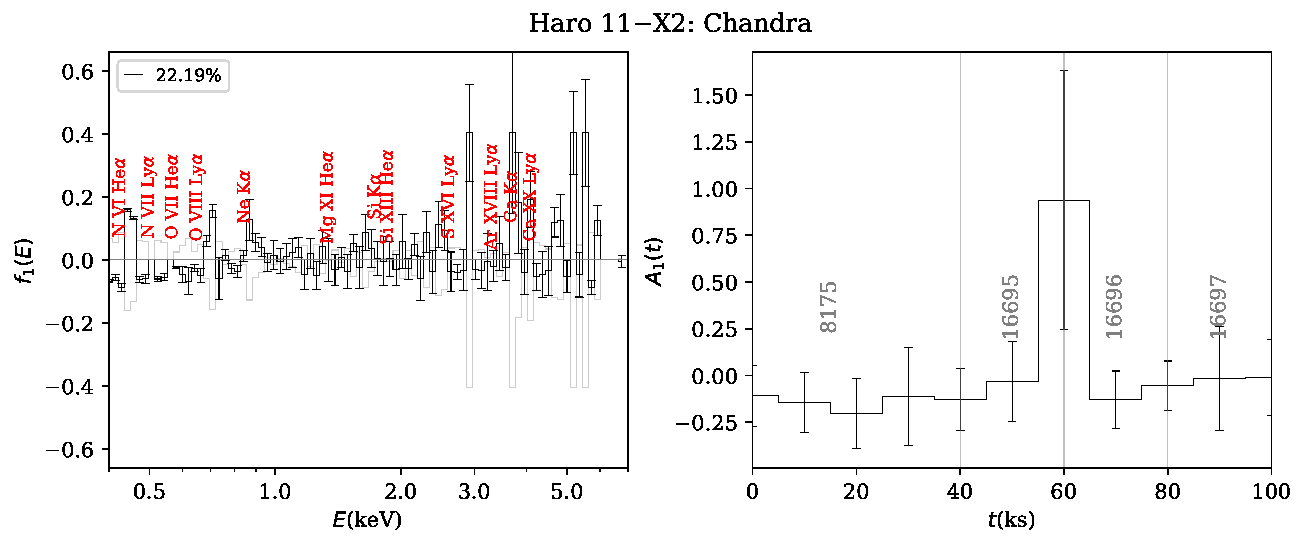}%
\end{center}
\caption{
Normalized PCA components $f_{k}(E)$ (left panels) of the \textit{XMM-Newton} pn+MOS2 (top), \textit{Chandra} ACIS-S X1 (middle), and X2 (bottom) of Haro\,11, along with the corresponding time-binned light curves $A_{k}(t)$ (right). The fractional percentage variability of each PCA component is also given in the left panel. 
The energies corresponding to the possible lines that arise from H-like and He-like ions are marked.
\label{haro11:fig:pca:1}
}
\end{figure*}

Principal component analysis (PCA) is a commonly used eigenvector-based approach for conducting multivariate statistics, which has been described in several books \citep[e.g.,][]{Jolliffe2002}. It can be employed to disentangle different spectral features primarily contributing to the intricate source variability by decomposing the time-stacked data into a set of PCA components that are independent and variable, thus providing the degrees of variability with fractional contribution orders. It has been widely used in the field of astronomy as a valuable tool for multivariate research. Its application can be traced back to some works on the stellar spectral classification \citep{Deeming1964,Whitney1983}, followed by spectral studies of galaxies \citep{Faber1973,Bujarrabal1981,Efstathiou1984}, and UV data analyses of quasars \citep{Mittaz1990,Francis1992}, in addition to imaging spectroscopy of the ISM \citep{Heyer1997,Brunt2009}. This approach has also been extensively used for spectral analyses of high-mass X-ray binaries \citep{Malzac2006,Koljonen2013}, AGNs \citep{Miller2008,Parker2014,Gallo2015,Parker2017,Parker2018}, and blazar variability \citep{Gallant2018}. 

PCA is a statistical technique that can split timing spectroscopic observations into sets of orthogonal spectral and temporal eigenvectors together with ordered eigenvalues of time variability. The process of projecting the time-stacked spectra onto the eigenvectors results in a series of diagnostic eigenvalues, which are associated with the variability contributions of ordered spectral sources resulting in the entire spectrum. We denote each paired decomposed spectral source and its time series as a principal component, identified by the order $i = 1, 2, ..., n$, where $n$ corresponds to the number of time slices in each dataset binned at $m$ energy channels. The variability fractions in the data are described by the normalized eigenvalues. The determination of spectral components and their light curves involves the eigen-decomposition of an $m \times n $ matrix, containing spectroscopic data sliced at $n$ time intervals and binned at $m$ energy channels.

For our eigenvector multivariate analysis, we used a modified version of the \textsc{pca} Python code developed by \citet{Parker2015} using the singular value decomposition \citep[SVD;][]{Press1997} function from the \textsf{NumPy} linear algebra \citep{Harris2020}. This program transfers the given $n$ time-segmented PHA data into a rectangular array ($m \times n $) consisting of a series of $n$ spectral data binned within $m$ energy channels, which is then decomposed using the SVD function into a 2D array ($m \times m$) including PCA components $F_{m}(E)$ sorted by their contribution orders, an $n$-array of eigenvalues associated with the variance of each PCA component, and a square matrix ($n \times n$) containing eigenvectors providing the time-binned PCA light curves $A_{n}(t)$. The PCA components and light curves represent the spectral shape and temporal change of the variable features present in a complex source, respectively. The determination of the fractional variability of each PCA component is achieved by dividing the corresponding eigenvalue by the sum of all eigenvalues. The method employed for calculating errors in the PCA program is based on the given spectra undergo random perturbations and SVD recalculations \citep[see][]{Miller2007}.

To conduct PCA, we employed the reprocessed event of each \textit{XMM-Newton} dataset to create the time-filtered events at time intervals of 10\,ks using the \textsc{sas} function \textsf{evselect}. The time-sliced event data similarly include the suitable patterned events and the PI channels appropriate for EPIC-pn and MOS2, in addition to the exclusion of the flaring background-affected events. The \textsf{especget} tool was utilized to generate time-stacked PHA files from the time-filtered events using the same source and background circular regions described in Section~\ref{haro11:observation}. For the \textit{Chandra} ACIS-S observations, the event files were filtered at time stacks of 10\,ks using the \textsc{ciao} task \textsf{dmcopy}. The source and background spectra, as well as the corresponding response data, were made from the time-filtered events with the \textsc{ciao} task \textsf{specextract}, set to correct point-like PSF in the ARF data. After the data reduction, we proceeded with supplying the time-sliced flux data of the source and background to the PCA program, in addition to the effective area data, which yielded the PCA components, time series, and normalized eigenvalues for each dataset. The effective area column $A_{\rm eff}(E)$ from the ARF was used to transfer the count value $C(E)$ in each energy bin $E$ to flux one $F(E)$ according to $F(E)= C(E) / (A_{\rm eff}(E) t)$, where $t$ is the exposure time. For each dataset, a mean spectrum, $F_{\rm mean}(E)$ was derived from the background-corrected PCA spectra, which was used to obtain the normalized PCA spectra as follows $f_{k}(E) = [F_{k}(E) - F_{\rm mean}(E)]/F_{\rm mean}(E)$.

Figure~\ref{haro11:fig:pca:3} presents the eigenvalue-fraction diagrams the \textit{XMM-Newton} combined pn and MOS2, as well as the \textit{Chandra} ACIS-S X1 and X2, in which the normalized eigenvalues of the principal components are plotted against the corresponding eigenvector orders. The number of nonzero eigenvalues is proportional to the number ($n$) of time-segmented PHA data in each dataset. These diagrams can be used to ascertain the statistical significance of the PCA components. The line in each diagram depicts the linear regression between logarithmic values of fractional variability and eigenvector orders, starting from three onward, which allows us to distinguish significant PCA components from those possibly created by noise fluctuations \citep[see, e.g.,][]{Koljonen2013,Parker2015}. Accordingly, the first two components in the \textit{XMM-Newton} pn+MOS2 data and the \textit{Chandra} X1, and the first component in \textit{Chandra} X2 are expected to be statistically significant based on the eigenvalue-fraction diagrams.

Figure~\ref{haro11:fig:pca:1} shows the PCA components and time series of the \textit{XMM-Newton} pn+MOS2, as well as the \textit{Chandra} ACIS-S X1 and X2. A series of peaks and troughs are seen in the principal components, which might correspond to variable emission and/or absorption features, possibly from ionized powerful superwinds. We see that these features are located at energies, where we could expect lines from H-like and He-like ions of different elements, such as \ionic{S}{xv} He$\alpha$ 2.45\,keV and \ionic{Ca}{xx} Ly$\alpha$ 4.11\,keV as labeled in the plots. The respective light curves imply that the principal components are primarily linked to some brightness incidents happening during the observations (see Fig.\,\ref{haro11:fig:lc}). The orders higher than 2 are not statistically significant in XMM and X1, since they follow the linear regression line in Figure~\ref{haro11:fig:pca:3}, so they do not seem to represent any spectral feature. We see that the second time series $A_{2}(t)$ of the \textit{Chandra} X1 partially follows the trend in $A_{1}(t)$, so it might be associated with the features of a superwind. We should note that X1 could also contain a wide-angled bipolar outflow based on observations made with the Fibre Large Array Multi Element Spectrograph (FLAMES) on the VLT \citep{James2013}, which can explain the double spectral features of an outflow in PCA. Especially, the time series $A_{2}(t)$  of X1 also has a peak that is co-temporal with the peak in $A_{1}(t)$. This means that they could be related to some brightening incidents that occurred at the beginning of the 2006 observation.

Previously, \citet{Parker2017} exploited PCA to reexamine the existence of outflowing absorption lines in the X-ray variability of a narrow-line Seyfert 1 galaxy. As labeled in Fig.\,\ref{haro11:fig:pca:1}, the principal spectra may contain several emission and/or absorption lines from H-like and He-like ions of various elements, which could originate from the outflow similar to those identified in the UV observations of Haro\,11 \citep{Grimes2007,Hayes2016,Oestlin2021,Komarova2024}. Especially, some hydrodynamic simulations of starburst superwinds also predicted the formation of radiatively cooling expanding regions along with hot ($10^{6-8}$\,K), X-ray-emitting layers, where the UV lines from highly ionized ionic species could be generated \citep[][]{Wuensch2017,Danehkar2021,Danehkar2022}. Recent hydrodynamic simulations conducted by \citet{Yu2021} demonstrated that H-like and He-like K$\alpha$ emission lines from O, Mg, Si, S, Ar, Ca, and Fe ions could be produced in the X-rays of starburst outflows. Moreover, several H-like and He-like emission lines of O, Ne, Mg, Si, S, Ar, and Fe were detected in the X-ray observations of starburst-driven outflows in NGC\,253 \citep{Ptak1997,Bauer2007,Bauer2008,Mitsuishi2013,Lopez2023} and M82 \citep{Ptak1997,Ranalli2008,Liu2011,Lopez2020}. However, the available low-count X-ray data of Haro\,11 do not allow us to confirm the presence of H-like and He-like emission lines through spectral modeling. 

\subsection{Spectral analysis}
\label{haro11:spec}

\begin{figure}
\begin{center}
\includegraphics[width=0.45\textwidth, trim = 0 0 0 0, clip, angle=0]{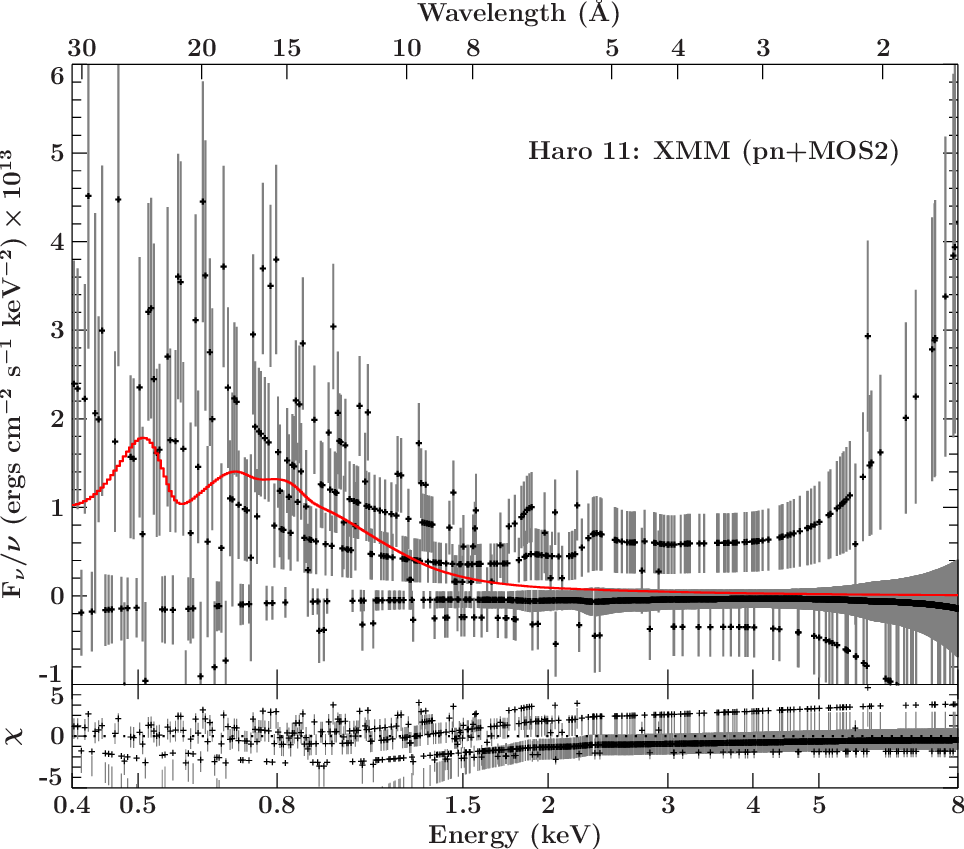}\\
\includegraphics[width=0.45\textwidth, trim = 0 0 0 0, clip, angle=0]{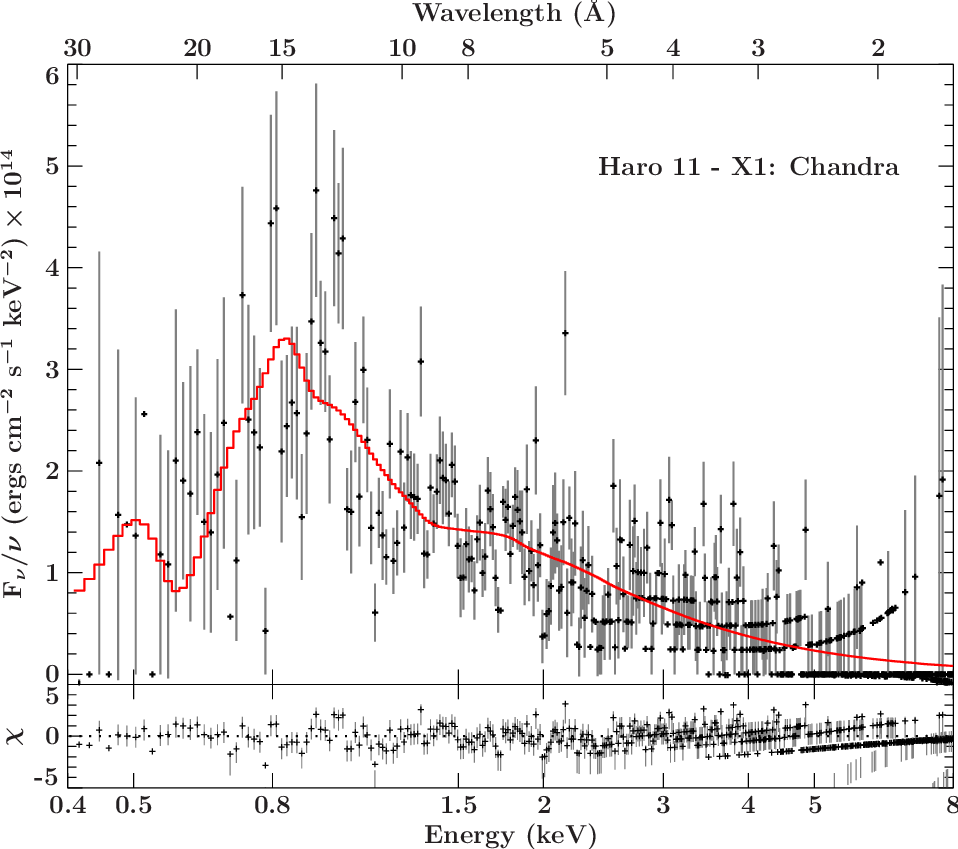}\\
\includegraphics[width=0.45\textwidth, trim = 0 0 0 0, clip, angle=0]{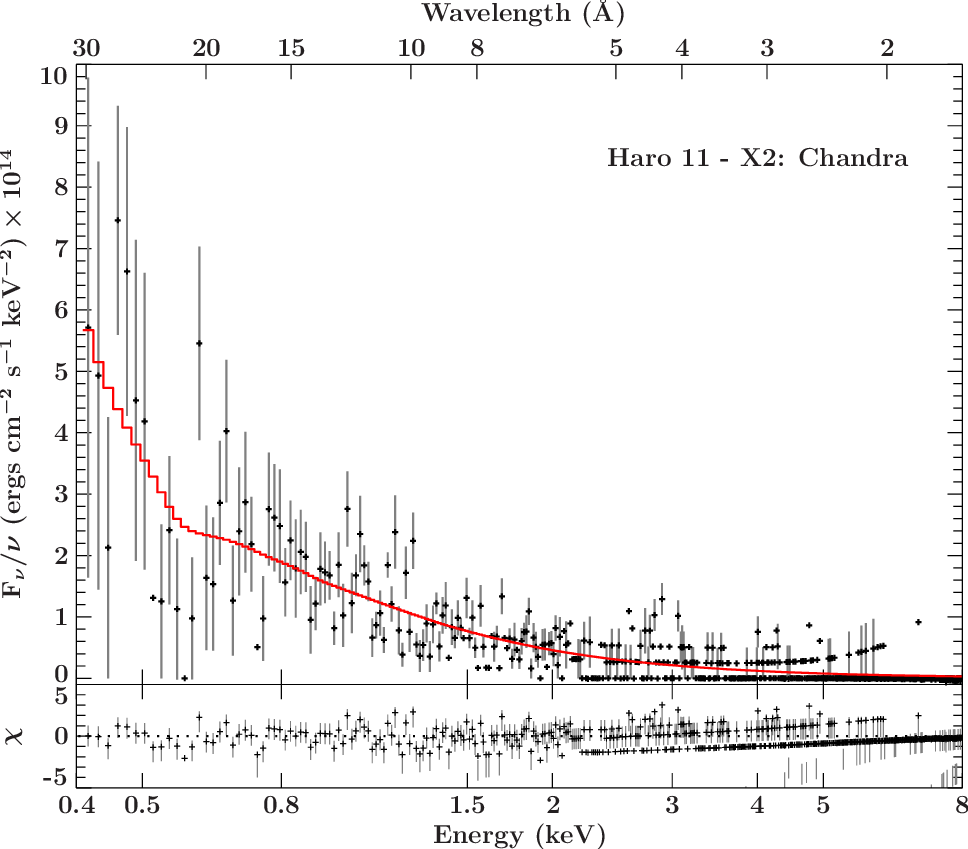}%
\end{center}
\caption{\textit{XMM-Newton} EPIC-pn+MOS2 (top), \textit{Chandra} ACIS-S X1 (middle), and X2 (bottom) observations of Haro\,11, along with the spectral models \textsf{constant}\,$\times$\,\textsf{phabs}\,$\times$\,(\textsf{diskbb}\,$+$\,\textsf{zpowerlw}) and \textsf{constant}\,$\times$\,\textsf{phabs}\,$\times$\,(\textsf{zbbody}\,$+$\,\textsf{zpowerlw}) (in red) fitted to the \textit{XMM-Newton} and \textit{Chandra} data, respectively. 
\label{haro11:fig:spec}
}
\end{figure}

\begin{table*}
%\centering
\begin{center}
\caption{Best-fitting parameters for the \textsc{xspec} models of the \textit{XMM-Newton} EPIC-pn+MOS2, and \textit{Chandra}/ACIS-S X1 and X2 data.  
\label{haro11:model:param}
}
\begin{tabular}{lcccc}
\hline\hline
\noalign{\smallskip}
{\textsc{xspec}} & {Parameter} & {XMM pn+M2}   & {CXO X1}  & {CXO X2} \\
\noalign{\smallskip}
\hline 
\noalign{\smallskip}
&  R.A. (J2000) \dotfill &  00:36:52.43 & 00:36:52.4278  & 00:36:52.6981 \\
\noalign{\smallskip}   
&  Decl. (J2000) \dotfill &  $-$33:33:16.69 &  $-$33:33:16.694 & $-$33:33:17.166 \\
\noalign{\smallskip}  
&  Extracted Source Radius ($''$)\ldots \dotfill &  36 & 1.6  & 1.1 \\ 
\noalign{\smallskip}
& Source Counts\,$^{\rm \bf a}$  \dotfill &  $ 968$ & $1330$ & $627$ \\ 
\noalign{\bigskip}
\textsf{phabs} & $N_{\rm H}(10^{21}\mathrm{cm}^{-2})$ \dotfill & $ {  4.29 }_{ -1.72 }^{ + 0.69 } $ & $ { 11.46 }_{ -1.02 }^{ + 1.75 } $ & $ {  1.16 }_{ -0.70 }^{ + 0.84 } $     \\ 
\noalign{\medskip}
\textsf{diskbb}  & $kT_{\rm in}$(eV) \dotfill  & $ { 161.3 }_{ -14.7 }^{ +50.8 } $   & \ldots  & \ldots \\
\noalign{\smallskip}  
                  & $K_{\rm dbb}$($10^{2}$)\,$^{\rm \bf b}$  \dotfill  &  $ {  6.46 }_{ -6.02 }^{ + 3.54 } $   &  \ldots & \ldots \\   
\noalign{\medskip}
\textsf{zbbody}   & $kT$(eV) \dotfill  & \ldots &  $ { 84.3 }_{ -7.4 }^{ + 8.3 } $ & $ { 29.5 }_{ -29.4 }^{ +11.0 } $ \\ 
\noalign{\smallskip}  
                  & $K_{\rm bb}$($10^{-6}$)\,$^{\rm \bf c}$  \dotfill  & \ldots &  $ { 138.31 }_{ -121.02 }^{ +392.50 } $ &  $ { 38.53 }_{ -38.53 }^{+\cdots } $ \\   
\noalign{\medskip}           
\textsf{zpowerlw} & $\Gamma$ \dotfill  & $ {  1.82 }_{ -0.65 }^{ + 0.77 }$    &   $ {  2.28 }_{ -0.15 }^{ + 0.17 } $ &   $ {  1.98 }_{ -0.22 }^{ + 0.21 } $ \\ 
\noalign{\smallskip}
                   & $K_{\rm p}$($10^{-5} \frac{\mathrm{photons}}{\mathrm{keV\,cm^{2}\,s}}$) \dotfill & $ { 12.92 }_{ -6.44 }^{ +11.78 } $    &   ${  3.45 }_{ -2.55 }^{ +11.53 } $ &  $ { 0.76 }_{ -0.76 }^{ +1.69 } $ \\
\noalign{\medskip} 
& C-stat/d.o.f. \dotfill &  $ 1.02$ & $0.95$ & $0.82$ \\ 
\noalign{\bigskip}
                   & $F_{\rm X}$($10^{-14} \mathrm{ergs}\,\mathrm{s}^{-1}\,\mathrm{cm}^{-2}$)\,$^{\rm \bf d}$ \dotfill & $    15.01^{+13.69}_{-7.48}$   &  $    10.99^{+36.73}_{-8.12}$ &  $    4.78^{+10.64}_{-4.78}$ \\
\noalign{\smallskip}         
                   & $L_{\rm X}$($10^{40} \mathrm{ergs}\,\mathrm{s}^{-1}$)\,$^{\rm \bf d}$ \dotfill & $    12.08^{+11.01}_{-6.02}$    &  $    8.84^{+29.55}_{-6.53}$ &  $    3.84^{+8.56}_{-3.84}$ \\
\noalign{\smallskip}
\hline
\noalign{\smallskip}
\end{tabular}
\end{center}
\begin{tablenotes}
\item[1]\textbf{Notes.} The \textsc{xspec} models \textsf{zbbody} and \textsf{zpowerlw} are in the rest frame ($z=0.020598$). 
$^{\rm \bf a}$ Source counts over 0.4--10\,keV for \textit{XMM-Newton} and over 0.4--8\,keV for \textit{Chandra}.
$^{\rm \bf b}$~Multi-temperature disk blackbody normalization defined as $f^{-4} (R_{\rm in}/D_{10})^2 \cos i$, 
where $R_{\rm in}$ is the inner radius (unit in km) of the disk, $D_{10}$ the distance to the source in units of 10\,kpc, $f$ the spectral hardening factor, and $i$ the inclination of the disk. 
$^{\rm \bf c}$~Blackbody normalization defined as $L_{39}/D_{10}^2$, where $L_{39}$ is the source luminosity ($10^{39}$\,erg\,s$^{-1}$), and $D_{10}$ the source distance (10\,kpc). 
$^{\rm \bf d}$~The unabsorbed X-ray flux and luminosity ($D=82$\,Mpc) over the 0.4--8 keV energy band. 
\end{tablenotes}
\end{table*}

To further evaluate the X-ray features of Haro\,11, we used phenomenological models to reproduce the curvatures in the \textit{XMM-Newton} and \textit{Chandra} observations. To improve the fitting statistics, we modeled the soft excess ($<1.1$\,keV) using a (multi-)blackbody accretion disk spectrum, which might originate from either an AGN or BHB in X1, and the hard band ($>1.1$\,keV) using a power-law model, in addition to a line-of-sight absorbing column. Our spectral analysis was conducted using the Interactive Spectral Interpretation System \citep[\textsc{isis} v\,1.6.2-51;][]{Houck2000}, which provides access to several S-Lang libraries, as well as the spectral models included in the X-ray spectral fitting package \textsc{xspec} \citep[v\,12.13.0;][]{Arnaud1996}  of the HEAsoft collection (v\,6.31.1). The Simulated Annealing (SA) minimization algorithm \citep{Corana1987}, together with the ``Cash'' statistic \citep[C-stat;][]{Cash1979}, were used to model simultaneously all the background-subtracted \textit{Chandra}/ACIS-S data. While the C-stat is the most effective method for the analysis of low-count data, the SA optimization method is also promising for extremely low counts, albeit with a highly intensive use of computing resources. We also simultaneously modeled the source and background data for each of the \textit{XMM-Newton} dataset using the C-stat and SA method, with the background spectrum fitted by two broken power-law components. The source spectra, along with the corresponding spectral models, are shown in Figure~\ref{haro11:fig:spec}.

For the soft excess in the XMM data, a multicolor disk blackbody model \citep[\textsf{diskbb}; see, e.g.,][]{Mitsuda1984,Zimmerman2005} was used, since they include both the X-ray sources X1 and X2. However, a simple blackbody model (\textsf{zbbody}) in the rest frame ($z=0.020598$) was sufficient to describe the soft excesses of X1 and X2 observed by the \textit{Chandra} ACIS-S. The power-law model (\textsf{zpowerlw}) that reproduces the hard excess has the two free parameters: photon index ($\Gamma$) and normalization factor ($K_{p}$). To recreate the drop (absorbed) pattern seen in the soft excess, an absorption component (\textsf{phabs}) was applied to both the soft and hard components, yielding the hydrogen column density ($N_{\rm H}$) of line-of-sight absorbing material. The spectral models, which were fitted to the 0.4--10\,keV energy range of the XMM data and 0.4--8\,keV of CXO as seen in Fig.~\ref{haro11:fig:spec}, are as follows: \textsf{constant} $\times$\,\textsf{phabs}\,$\times$\,(\textsf{diskbb}\,$+$\,\textsf{zpowerlw}) for XMM and  \textsf{constant} $\times$\,\textsf{phabs}\,$\times$\,(\textsf{zbbody}\,$+$\,\textsf{zpowerlw}) for CXO. The cross-normalization \textsf{constant} was set to the index number of each active dataset,\footnote{This is done via the use of \textsf{constant(Isis\_Active\_Dataset)} in \textsc{isis}.} which allows data-dependent modeling of different epochs and instruments.

The best-fitting parameters of the spectral models are listed in Table~\ref{haro11:model:param}. The \textsc{isis} standard procedure for confidence limits (\textsf{conf\_loop}) was used to determine the best-fitting parametric values and the corresponding uncertainties at 90\% confidence. 
One can note that the \textit{XMM-Newton} data were slightly different ($\Gamma \sim 1.8$)
than the \textit{Chandra} data, albeit consistent based on the confidence limits. 
The comparison of \textit{Chandra}/ACIS-S and \textit{XMM-Newton} model parameters indicates that the source X1 in the \textit{Chandra} is more than 2 times more obscured than what is seen in the XMM, while the column density along the line of sight toward X2 is 10 times lower than that in X1. 
Figure~\ref{haro11:fig:spec} also shows that the soft excess ($<1$\,kev) in X2 is stronger than X1, which is suggestive of a much lower absorbing material in X2. Interestingly, the absorbing column for the \textit{XMM-Newton} spectra, which are made up of both X1 and X2, are lower than that derived for X1 with the \textit{Chandra}.
The blackbody temperature and normalization in X2 are lower than those in X1. This means that the soft excess in the \textit{XMM-Newton} data mostly originates from X1, where it is subject to the possible accretion associated with an AGN or a BHB. The power-law model describing the hard excess in the source X2 ($\Gamma \sim 2$) is slightly harder than that derived for X1 ($\Gamma\sim 2.3$). However, the power-law normalization scale for the X1 source is much higher than that in X2, meaning that X1 is also brighter in the hard excess (see Table~\ref{haro11:model:param}). 

The previous study of the 2006 \textit{Chandra} observation by \citet{Prestwich2015} using the aperture of $1.1''$ also indicated that X2 is much softer and slightly less obscured than X1, while X1 is twice as luminous as X2, which is similar to our finding in Table~\ref{haro11:model:param}. Moreover, a recent analysis of the stacked \textit{Chandra}/ACIS-S data by \citet{Gross2021} with the extraction radius of $1.7''$ implied that X1 could also have an absorbing column that is four times higher than that of X2. The power-law photon index ($\Gamma\sim 2$) we derived for X2 is in reasonable agreement with that found by \citet{Gross2021}, though our best-fitting parameter values for X1 are slightly different. This discrepancy could be explained by our approach of simultaneously fitting a spectral model to all the spectra rather than using a stacked spectrum. 

In Table~\ref{haro11:model:param}, we also list the unabsorbed X-ray flux ($F_{\rm X}$) and the X-ray luminosity ($L_{\rm X}$) at $D=82$\,Mpc. The flux of each model, in addition to the luminosity,  were derived over the energy range 0.4 to 8 keV using an S-Lang flux calculation routine (\textsf{model\_flux}) from the custom \textsc{isis} functions (distributed by M.\,Nowak).\footnote{\url{https://space.mit.edu/home/mnowak/isis_vs_xspec/}} 
The confidence levels of the X-ray fluxes and luminosities are based on the those derived from the power-law normalization factors. The high uncertainties could be related to the X-ray variability and low counts. One can note that the source X1 is twice as bright as X2. We also see that the X-ray luminosity in the \textit{XMM-Newton} EPIC-pn is roughly equal to the luminosity sum of the sources X1 and X2 collected by the \textit{Chandra} ACIS-S.
Our X-ray luminosity estimation for the source X2 ($L_{\rm X} \sim 4 \times 10^{40} $\,erg\,s$^{-1}$) agrees approximately with the ULX luminosity ($L_{\rm X}\sim 5$--$6 \times 10^{40}$\,erg s$^{-1}$) derived by \citet{Gross2021}, albeit with $D=84$\,Mpc. Moreover, we obtained an X-ray luminosity for X1 ($L_{\rm X} \sim 9 \times 10^{40}$\,erg\,s$^{-1}$) similar to the extremely hard X-ray luminosity ($L_{\rm X}\sim 10^{41}$\,erg s$^{-1}$) found previously \citep[][]{Prestwich2015,Gross2021}.

\begin{figure}
\begin{center}
\includegraphics[width=0.45\textwidth, trim = 0 0 0 0, clip, angle=0]{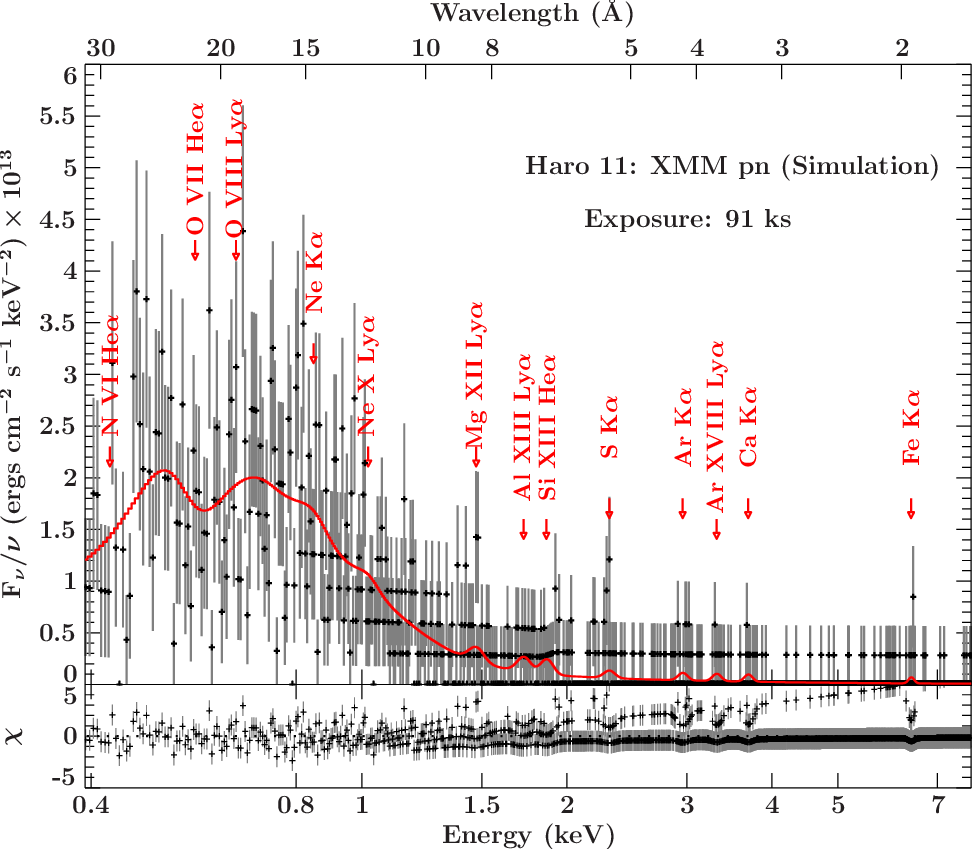}\\
\includegraphics[width=0.45\textwidth, trim = 0 0 0 0, clip, angle=0]{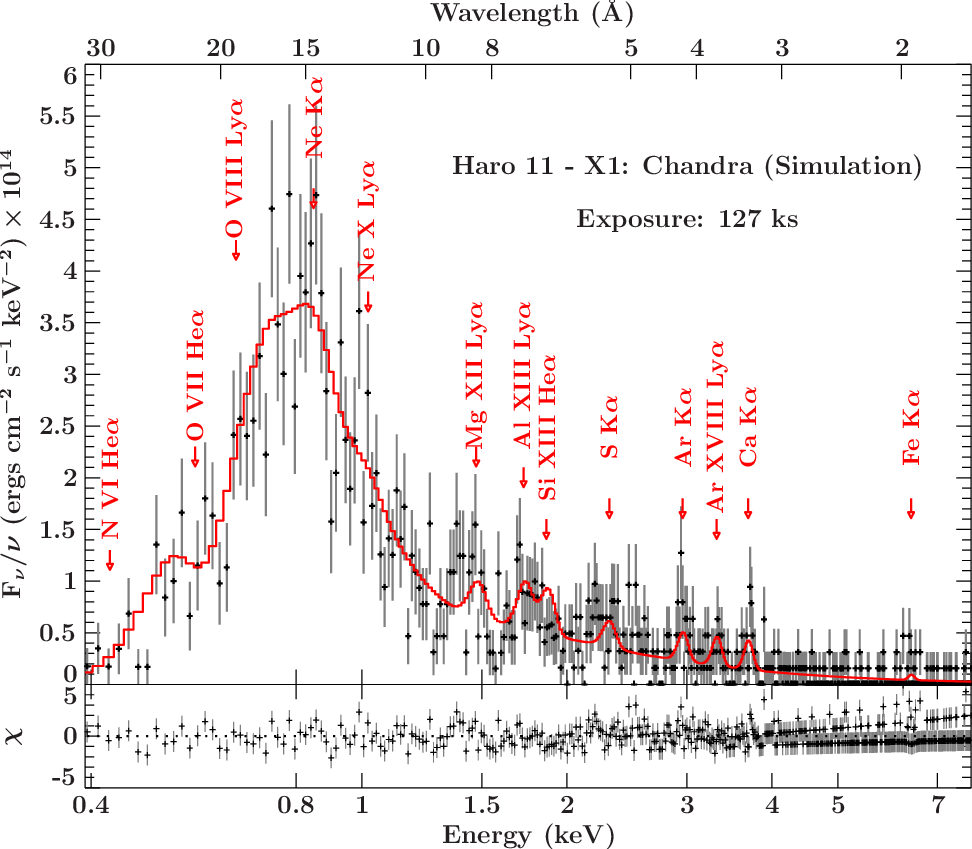}\\
\includegraphics[width=0.45\textwidth, trim = 0 0 0 0, clip, angle=0]{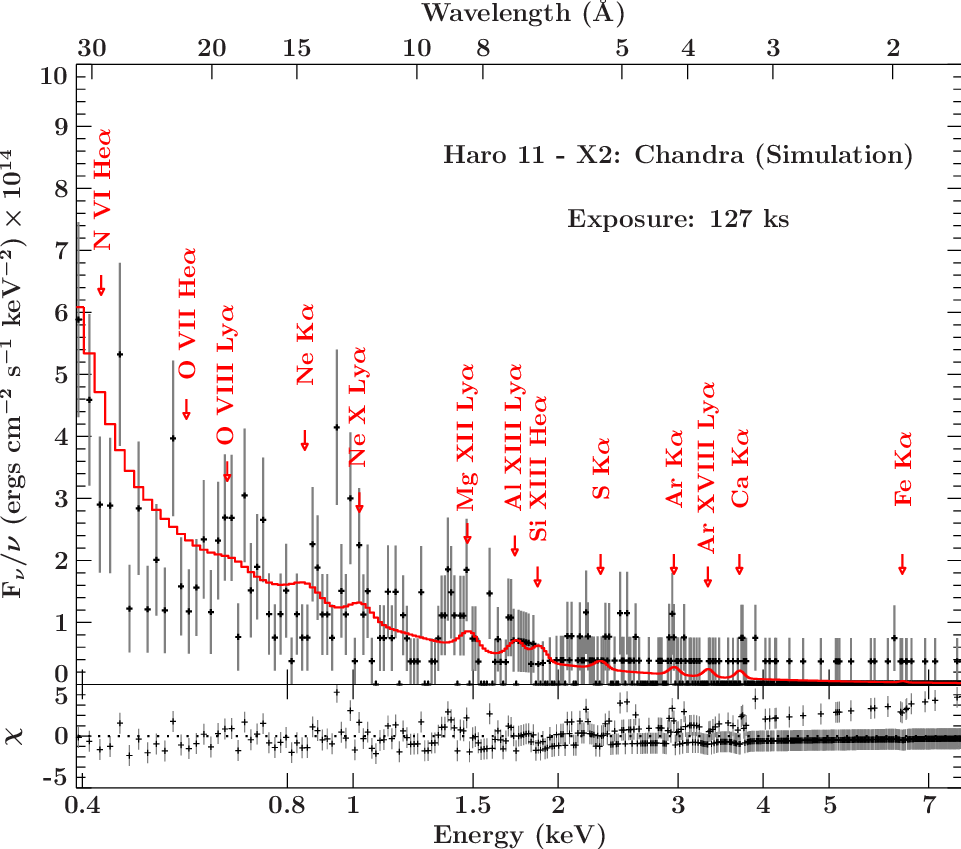}%
\end{center}
\caption{Simulated spectra of the \textit{XMM-Newton} EPIC-pn (91\,ks) and \textit{Chandra} ACIS-S X1 and X2 sources (127\,ks) using the model parameters listed in Table~\ref{haro11:model:param}, including the emission lines suggested by our PCA, as labeled in Fig.\,\ref{haro11:fig:pca:1}.
\label{haro11:fig:sim}
}
\end{figure}

\subsection{Simulations}
\label{haro11:sim}

\begin{figure*}
\begin{center}
\includegraphics[width=0.75\textwidth, trim = 0 0 0 0, clip, angle=0]{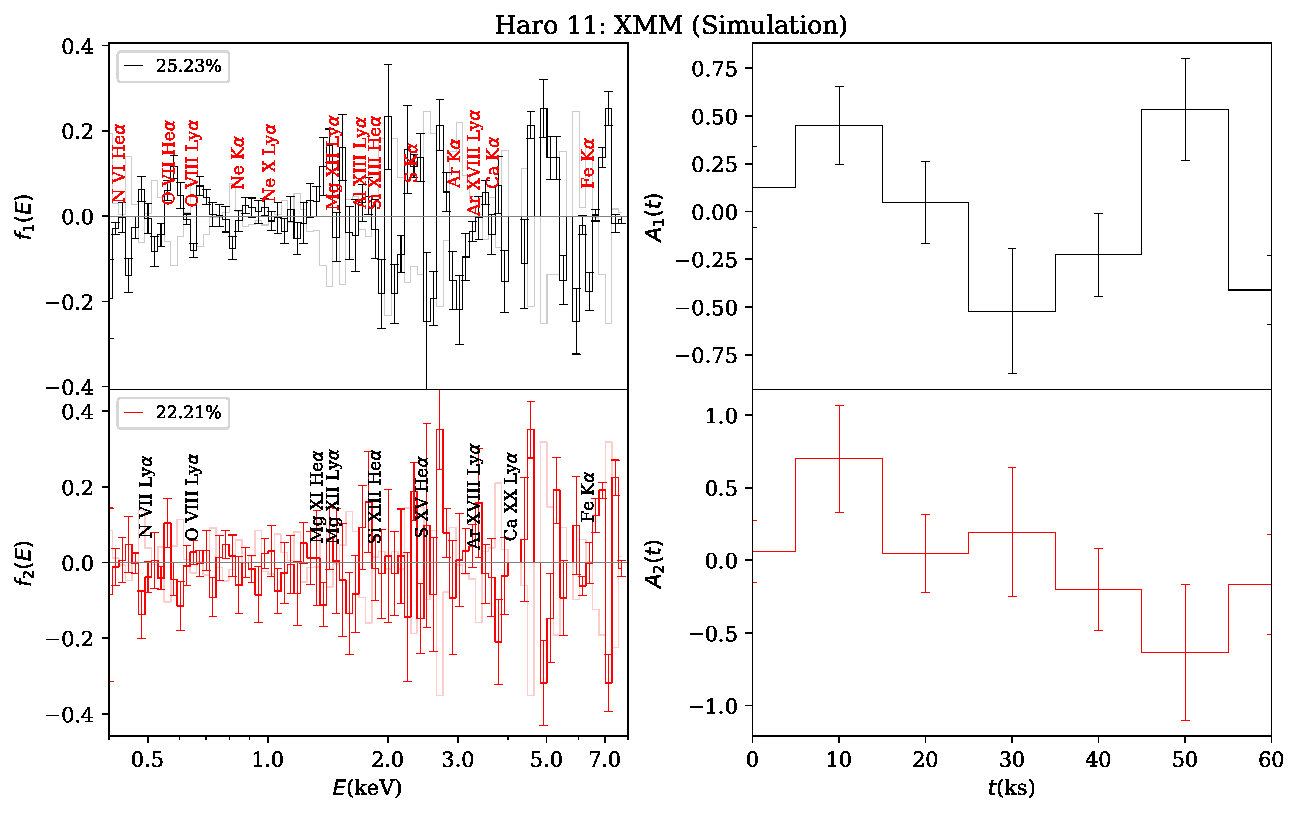}
\end{center}
\caption{Normalized PCA components $f_{k}(E)$ and the corresponding time series $A_{k}(t)$ of the simulated \textit{XMM-Newton} EPIC-pn spectra segmented at 10\,ks with the model parameters listed in Table~\ref{haro11:model:param}, including the \textsf{zgauss} models of the emission lines given in Table~\ref{haro11:model:param} with variable normalization factors as described in text.
\label{haro11:fig:pca:sim}
}
\end{figure*}

To assess the feasibility of PCA-suggested lines, we simulated spectra for \textit{XMM-Newton}/EPIC-pn and \textit{Chandra}/ACIS-S by adopting the total exposures of all combined \textit{XMM-Newton} and \textit{Chandra} data, respectively. Our simulations were conducted with the \textsc{isis} function \textsf{fakeit} using the RMF and ARF of the \textit{XMM-Newton}/EPIC-pn in 2004 and the \textit{Chandra}/ACIS-S data collected in 2006. We used the same spectral models of Sect. \ref{haro11:spec}, but included the  \textsc{xspec} models \textsf{zgauss} to simulate possible features of emission lines, namely \textsf{constant}\,$\times$\,\textsf{phabs}\,$\times$\,(\textsf{diskbb} $+$ \textsf{zpowerlw} $+$ $\sum$\,\textsf{zgauss}) for XMM-pn and  \textsf{constant}\,$\times$\,\textsf{phabs}\,$\times$\,(\textsf{zbbody} $+$ \textsf{zpowerlw} $+$ $\sum$\,\textsf{zgauss}) for CXO-X1/X2 (see Fig.\,\ref{haro11:fig:sim}). We set the model parameters to those values determined from our spectral analysis. The energy-independent factor in the \textsc{xspec} component \textsf{constant} of each spectral model was adjusted to reproduce exactly the same source count given in Table~\ref{haro11:model:param}. We shifted the \textsf{zgauss} components with an outflow velocity of $v_{\rm out}=-290$ km\,s$^{-1}$ in the rest frame according to the absorption lines blueshifted with 200--$380$\,km\,s$^{-1}$ detected in the UV observations collected with the Far Ultraviolet Spectroscopic Explorer (FUSE) and the Cosmic Origins Spectrograph (COS) on the HST \citep{Grimes2007,Hayes2016}, and assumed a Gaussian RMS width of $\sigma=10$ eV.

The normalization factors ($K$) of the \textsf{zgauss} functions were iteratively modified to reproduce similar features seen in the observed spectra. Table~\ref{haro11:sim:param} presents the aforementioned factors adopted for the \textsf{zgauss} models in the simulated spectra, which may be served as the upper limits to the emission lines possibly presented in the actual observations. Although it is difficult to distinguish any lines below 1\,keV and above 3\,keV,
our simulated spectra suggest the possible presence of the blueshifted emission lines \ionic{Mg}{xii} Ly$\alpha$ 1.471\,keV, \ionic{Al}{xiii} Ly$\alpha$ 1.727\,keV, \ionic{Si}{xiii} He$\alpha$ 1.865\,keV, and S K$\alpha$ 2.308\,keV in the XMM and CXO-X1 data as seen in Figure\,\ref{haro11:fig:sim}. In the case of CXO-X2 with extremely low photons, some features seen in the observation resemble the blueshifted lines \ionic{Mg}{xii} Ly$\alpha$ 1.471\,keV and \ionic{Si}{xiii} He$\alpha$ 1.865\,keV.
Our simulations indicate that the aforementioned emission lines in the medium band (1.1--2.6\,keV), if they are genuinely present, can be constrained using observations taken with an exposure of at least 300\,ks, which result in 3209, 3119, and 1034 counts in XMM-pn (0.4--10\,keV), CXO-X1 and -X2 (0.4--8\,keV), respectively.

%\vfill\null
%\vfill\eject
%\vfill\break

\begin{table}
%\centering
\begin{center}
\caption{Normalization factors ($K$) of the \textsc{xspec} model \textsf{zgauss} used to simulate possible emission lines in \textit{XMM-Newton}/EPIC-pn and \textit{Chandra}/ACIS-S X1 and X2 data in Fig.\,\ref{haro11:fig:sim}.  
\label{haro11:sim:param}
}
%\begin{tabular}{lC{4.1cm}cccc} % error with linenumbers in aastex631
\begin{tabular}{lcccc}
\hline\hline
\noalign{\smallskip}
Line & $E_l$(keV) & {XMM-pn} &   {CXO-X1}  & {CXO-X2}\\ 
\noalign{\smallskip}
\hline 
\noalign{\smallskip}
\ionic{N}{vi} He$\alpha$        & 0.426  & 8.0  & 1.0  & 0.2 \\
\ionic{O}{vii} He$\alpha$       & 0.569  & 8.0  & 1.0  & 0.2 \\
\ionic{O}{viii} Ly$\alpha$  & 0.653  & 8.0  & 1.0  & 0.2 \\
Ne K$\alpha$                            & 0.849  & 8.0  & 1.0  & 0.2 \\  
\ionic{Ne}{x} Ly$\alpha$        & 1.021  & 8.0  & 1.0  & 0.2 \\
\ionic{Mg}{xii} Ly$\alpha$  & 1.471  & 8.0  & 1.0  & 0.2 \\
\ionic{Al}{xiii} Ly$\alpha$ & 1.727  & 8.0  & 1.0  & 0.2 \\ 
\ionic{Si}{xiii} He$\alpha$ & 1.865  & 8.0  & 1.0  & 0.2 \\  
S K$\alpha$                             & 2.308  & 4.0  & 0.5  & 0.1 \\ 
Ar K$\alpha$                            & 2.958  & 4.0  & 0.5  & 0.1 \\ 
\ionic{Ar}{xviii} Ly$\alpha$ & 3.318  & 4.0  & 0.5  & 0.1 \\ 
Ca K$\alpha$                            & 3.691  & 4.0  & 0.5  & 0.1 \\ 
Fe K$\alpha$                            & 6.404  & 4.0  & 0.1  & 0.02 \\
\noalign{\smallskip}
\hline
\noalign{\smallskip}
\end{tabular}
\end{center}
\begin{tablenotes}
\item[1]\textbf{Notes.} The \textsf{zgauss} components are shifted with $v_{\rm out}=-290$ km\,s$^{-1}$ in the rest frame ($z=0.020598$) and have a line width ($\sigma$) of 10\,eV. The normalization factor unit is in $10^{-6}$\,photons\,cm$^{-2}$\,s$^{-1}$ in the line energy ($E_l$).
\end{tablenotes}
\end{table}

To see the effects of the line variability, we employed the same aforementioned spectral models to generate simulated 10\,ks-stacked \textit{XMM-Newton} spectra, but with moderately variable normalization factors in the model components. Our approach is similar to the method used by \citet{Parker2014a}, albeit with the Gaussian function heights tied to the amplitudes of the PCA time series extracted from the observations. To create the variable emission-line fluxes, we simply set the normalization factors of the \textsc{xspec} functions \textsf{zgauss} to $[1 + (A_{1}(t)+A_{2}(t))/2] \times K$, where $K$ is the value chosen from Table~\ref{haro11:model:param}, and $A_{1}(t)$ and $A_{2}(t)$ are the PCA time series derived from the \textit{XMM-Newton} observations plotted in Fig.\,\ref{haro11:fig:pca:1}. However, we caution that our assumption of uniform changes in the emission-line fluxes may not be physically correct due to the presence of different stratification ionization zones in an ionized outflow. To emulate the background noise, we added 2\% variation to the normalization factors of the \textsc{xspec} models \textsf{diskbb} and \textsf{zpowerlw} using variable random seeds with a normal distribution. The simulated spectra were created with the \textsc{isis} function \textsf{fakeit} with an interval exposure of 10\,ks using the RMF and ARF of the \textit{XMM-Newton}/EPIC-pn data, and were written to PHA files with the help of the Remeis \textsc{isis} functions (ISISscripts). 
We again loaded and decomposed them into principal components using the SVD function in the same Python program used for our PCA of the observations.

Figure~\ref{haro11:fig:pca:sim} shows the statistically significant PCA components derived from our simulated spectra with variable emission-line fluxes. It can be seen that variable emission-line fluxes in the spectral model produce a series of high and low points in the PCA spectra. Interestingly, the homogeneous variation in line fluxes, which could be linked to a varying ionization parameter of an ionized outflow, results in two PCA components roughly similar to those seen in Fig.\,\ref{haro11:fig:pca:1}. We should note that the features we see in the PCA components derived from the observations may not be associated with a single outflow. However, our simulation with a simple assumption supports the idea that varying fluxes in emission lines, likely from different stratification layers of superwinds, might be present in Haro\,11.

\section{Discussions}
\label{haro11:discussions}

\subsection{Multiphase ionized superwinds in Haro\,11}

We analyzed the X-ray variability of the starburst knots B (X1) and C (X2) of Haro\,11 using the hardness diagrams and eigenvector-based multivariate statistics. There are some hourly transient incidents in both sources seen in the hardness ratios, which may be associated with emission features from ionized superwinds according to the PCA components derived from the X-ray observations. 

The signature of ionized superwinds in Haro\,11 has been seen in UV observations. The possible X-ray outflows could be counterparts to the \ionic{O}{vi} $\lambda\lambda$1032,1038 emission found in HST imaging \citep{Hayes2016}, in addition to an \ionic{O}{vi}-absorbing component blueshifted with velocities of 200--$380$\,km\,s$^{-1}$ detected in the FUSE and HST/COS UV spectroscopic observations \citep{Grimes2007,Hayes2016}. Broad, low-ionization interstellar absorption lines were also seen in the FUSE observations \citep{Grimes2007}, which are often observed in UV observations of LBGs \citep{Shapley2003}. The blueshifted \ionic{O}{vi} absorption line implies the presence of ionized outflows similar to those seen in high-$z$ LBGs. The blueshifted UV absorption \ionic{Si}{ii} and \ionic{Si}{iv} lines were also identified in the FUSE data, suggesting the presence of multiphase outflows with velocities extended to 500\,km\,s$^{-1}$ within the neutral ISM \citep{Oestlin2021}. This can create shock-ionization, which may result in blueshifted emission lines in X-rays.

In addition, \citet{Grimes2007} speculated whether the \ionic{O}{vi} emission in Haro\,11 originates from radiative cooling or resonance scattering, while also suggested that a significant amount (20\%) of the supernova feedback could be lost owing to radiative cooling. Recently, some hydrodynamic simulations put forward the production of \ionic{C}{iv} $\lambda$1550 and \ionic{O}{vi} $\lambda$1037 emission via nonequilibrium ionization within radiatively cooling superwinds \citep{Gray2019,Danehkar2022}. Interestingly, Knot A in Mrk\,71, which is the nearest LyC emitter candidate, also exhibits diffuse \ionic{C}{iv} $\lambda$1550 emission with a remarkable link to catastrophic radiative cooling based on nonequilibrium computations \citep{Oey2023}. The detection of \ionic{C}{iv} associated with radiative cooling in Knot C (X2) in Haro\,11 requires deeper spatially resolved UV observations, in addition to high-resolution X-ray spectroscopy.

The evidence for superwinds in Haro\,11 is not limited to those seen in UV observations and is also visible in spatially resolved optical observations. Knot B (X1) seems to be the main source for fast outflows, with escape wind speeds ranging from 240 to 600\,km\,s$^{-1}$, according to integral field unit (IFU) observations made with the optical VLT/MUSE instrument \citep{Menacho2019}. An apparently broad component blueshifted at 400\,km\,s$^{-1}$ was also identified in the MUSE data, which was interpreted as ionized outflows away from knots B and C \citep{Sirressi2022}. Moreover, \citet{Sirressi2022} argued that Knot C is a depleted area formed by a previous outflow, as suggested by \citet{Menacho2019}, while the outflow in Knot B seems to be young and less developed. Previously, \citet{Hayes2007} also found evidence for a bipolar morphological outflow in Knot C based on the Ly$\alpha$ image. Moreover, a wide-angled bipolar outflow centered on Knot B was also proposed by \citet{James2013} based on optical IFU observations made with VLT/FLAMES. 

Numerical simulations demonstrated that the propagation of powerful outflows into the surrounding ISM results in multiphase stratified zones via collisional ionization, which leave observable imprints in X-rays \citep{Yu2021}, in addition to optical and UV lines \citep{Danehkar2021}. In particular, X-ray spectra of starburst-driven outflows in NGC\,253 and M82 showed several emission lines of various elements \citep{Liu2011,Mitsuishi2013,Lopez2020,Lopez2023}. As discussed by \citet{Menacho2019}, such a strong highly ionized outflow is capable of clearing galactic channels, which allows the escape of LyC radiation in Knot C (X2).

\subsection{Covering fractions and densities in X1 and X2}

Our spectral analysis of X1 and X2 based on the \textit{Chandra} ACIS-S data implies that X2 is less covered by the absorbing material than X1. The ISM around X2 could have lower column densities or some low-density channels, which would make it optically thin to the LyC radiation. In particular, the source X2, which is overall fainter in X-rays compared to X1, has a Ly$\alpha$ escape fraction of 6\%, whereas the escape fraction of Ly$\alpha$ in X1 was estimated to be $\lesssim 1$\% \citep{Oestlin2021}. This is consistent with the escape fractions of $3.4$\% and $5.1$\% in X1 and X2, respectively, determined from the leaking LyC luminosities in 903--912\,{\AA} \citep{Komarova2024}.

The narrow Ly$\alpha$ emission with no broad absorption recently seen with FUSE and HST/COS, along with the \ionic{Si}{ii} line profiles, hinted at a low ($\lesssim 50$\%) covering fraction of neutral gas in X2 \citep{Oestlin2021}. However, \citet{Rivera-Thorsen2017} deduced that the surrounding ISM around the knots is most likely either a clumpy density-bounded neutral medium or an ionization-bounded neutral medium within a highly ionized clumpy medium with an overall covering fraction of unity using the apparent optical depth method. %applied to the \ionic{Si}{ii} lines.
 They also suggested that the LyC escape might occur if the proportion of dense gas covering X2 is partly low enough. This necessitates that stellar feedback should significantly contribute to clearing pathways partially in X2 for ionizing radiation to penetrate the neutral medium. Previously, \citet{Sandberg2013} also concluded high optical depths but a low covering fraction ($\sim 10$\,\%) in all the knots through the spatially resolved observations of the sodium doublet (Na\,D) profiles.

Optical IFU spectroscopy of Haro\,11 with the VLT/MUSE studied by \citet{Menacho2021} revealed that the ISM around X2 is less dense ($\lesssim 35$\,cm$^{-3}$), hotter ($\gtrsim 16900$\,K), and less metallic ($12+\log{\,\rm O/H} \sim 7.46$) than X1  ($\sim 390$\,cm$^{-3}$, $\sim 9600$\,K, and $12+\log{\,\rm O/H} \sim 8.55$).
Using the IFU FLAMES on the VLT, \citet{James2013} also derived the same results: $n_{\rm e}=150$\,cm$^{-3}$, $T_{\rm e} = 16600$\,K, and $12+\log{\,\rm O/H} = 7.8$ for the ISM in Knot C (X2); and $n_{\rm e}=370$\,cm$^{-3}$, $T_{\rm e} = 12100$\,K, and $12+\log{\,\rm O/H} =8.25$ in in Knot B (X1). The LyC radiation and hot gaseous outflows are probably escaping through low-density channels in X2. Using the VLT/MUSE observations, \citet{Menacho2019} also found evidence for a fragmented, kiloparsec-scale superbubble shell around X2, which could be created by a past powerful superwind. This is supplemental to a low-density and high-temperature medium present in X2. Low-density channels and broken shells in the ISM surrounding X2, which might be created by powerful outbursts, could play a leading role in facilitating the escape of LyC and Ly$\alpha$ photons. In the future, if absorption lines are robustly constrained by high-resolution X-ray observations (e.g., XRISM), it will be possible to measure the outflow velocities in addition to the line equivalent widths (i.e., the column densities of ions produced in X-ray outflows).

%\subsection{X2: superwinds and LyC escape}
%\label{haro11:discussions:x2}

%\section{Discussions and Conclusion}
\section{Conclusion}
\label{haro11:conclusion}

% Move 1: Background information: original hypothesis
% Move 2: Summarizing and reporting key results: findings
% Move 3: Commenting on the key results: explanations for findings
% Move 4: Stating the limitation of the study: limitations
% Move 5: Making recommendations for the future implementation: need for further research

In this paper we have analyzed the 2004 and 2005 \textit{XMM-Newton} (EPIC-pn: 31.3\,ks and MOS2: 59.9\,ks)
and \textit{Chandra}/ACIS-S (127.3\,ks; see Table~\ref{haro11:obs:log}) observations of the Lyman break analog Haro\,11.
We used hardness-intensity diagrams to search for any possible spectral-state transition.
% in the spectral states, i.e., changing from soft to hard and faint to bright. 
We utilized principal component statistics to determine the spectral feature that is responsible for X-ray variability. We also conducted spectral modeling to reproduce the continua using a (multi-)blackbody spectrum and a power-law component covered by a line-of-sight absorbing column, in addition to the simulations of PCA-suggested lines. Our key findings are summarized as follows:

(i) Our timing analysis depicts some stochastically increases in the hardness ratios over the course of the observations (see Fig.~\ref{haro11:fig:lc}), which might be associated with the variability of emission lines made by ionized superwinds, as evidenced by the PCA light curves (Fig.~\ref{haro11:fig:pca:1}). Although the brightening incidents originate from both X1 and X2, the source X1 seems to contribute significantly higher counts to the X-ray variability.

(ii) Our statistical analysis indicates that the variability in the HR$_{2}$ hardness ratio of the XMM data is statistically significant according to all three normality tests (see Table~\ref{rtcru:stat:result}). In addition, the von Neumann statistics imply inhomogeneous changes in the $S + M$ light curves of the XMM observations. Moreover, the Lilliefors test shows marginally significant variations in the $S + M$ band of  X1.

(iii) Our eigenvector-based multivariate analyses of the \textit{XMM-Newton}/EPIC-pn+MOS2 data and the \textit{Chandra}/ACIS-S data of X1 and X2 indicate that the principal components responsible for the X-ray variability contain some peaks and valleys, which could be related to the emission lines from the starburst-driven outflow, similar to those detected in the UV observations \citep{Grimes2007,Hayes2016,Oestlin2021,Komarova2024}. Particularly, H-like and He-like emission lines of several elements, such as O, Mg, Si, S, Ar, and Fe, were detected in the X-ray spectra of starburst outflows in NGC\,253 \citep{Bauer2007,Mitsuishi2013,Lopez2023} and M82 \citep{Ranalli2008,Liu2011,Lopez2020}, and also predicted by hydrodynamic simulations of starburst galaxies \citep[][]{Yu2021}. The presence of highly ionized emission lines in Haro\,11 should be confirmed by deeper high-spectral-resolution X-ray spectroscopic observations in the future.  
Additionally, our PCA of the XMM and CXO-X1 data revealed a second principal component that has a deviation from the noise-related regression (see Fig.~\ref{haro11:fig:pca:3}). We see that the light curve $A_{2}(t)$ of X1 partially treads on the heels of $A_{1}(t)$ (see Fig.~\ref{haro11:fig:pca:1}), and both may have a connection to the same brightening incidents seen in the 2006 observation. This component might correspond to the spectral features of either another outflow or the opposite part of an extended bipolar outflow. Interestingly, \citet{James2013} suggested the presence of a wide-angled bipolar outflow in X1 based on spatially resolved IFU observations.
Alternatively, as our simulation suggests, an outflow with a varying ionizing source could also result in two PCA components (see Fig.\,\ref{haro11:fig:pca:sim}).

(iv) The time-averaged models of the \textit{Chandra}/ACIS-S spectra (see Table~\ref{haro11:model:param}) suggest that the absorbing column in X2 ($\sim 1.2\times 10^{21}$\,cm$^{-2}$) is much smaller than that in X1 ($\sim 11.5\times 10^{21}$\,cm$^{-2}$). 
Our findings indicate that the ISM around X2 (which has the higher LyC leakage) has either a lower covering fraction or a lower column density, which may be the result of a strong superwind.
Our spectral analysis also shows that X2 is slightly harder ($\Gamma \sim 2$) and twice fainter ($L_{\rm X} \sim 4 \times 10^{40} $\,erg\,s$^{-1}$) than X1 ($\Gamma \sim 2.3$ and $L_{\rm X} \sim 9 \times 10^{40} $\,erg\,s$^{-1}$). Moreover, our spectral modeling of the \textit{XMM-Newton} data, which includes both X1 and X2, supports the presence of absorbing material ($\sim 4.3\times 10^{21}$\,cm$^{-2}$). 

(v) The simulated spectra, based on our spectral models, suggest the presence of some blueshifted emission lines, such as \ionic{Mg}{xii} Ly$\alpha$ 1.471\,keV and \ionic{Si}{xiii} He$\alpha$ 1.865\,keV, in X1 and X2, which could originate from ionized superwinds driven by SSCs in Knots B and C. If these lines truly exist, they can only be constrained using observations much deeper than the currently available data. Our simulated spectra with varying emission-line fluxes, which could be produced by different ionization zones of an ionized superwind, also result in the PCA spectra resembling those derived from our analysis of the observations.

X-ray studies of Haro\,11 would greatly improve our understanding of the underlying physics behind the LyC and Ly$\alpha$ leakage in high-$z$ star-forming galaxies. In particular, radiation escaping from starburst galaxies in the early Universe is believed to significantly contribute to cosmic reionization \citep{Cowie2009,Robertson2015,Mitra2015}. Haro\,11 can serve as a prominent local analog of LyC emitters. The use of future high-spectral-resolution X-ray missions such as XRISM \citep{Tashiro2020} has the potential to provide further detailed information on the X-ray characteristics of Haro\,11 and the physical phenomena that facilitated the escape of ionizing radiation from starburst galaxies during the reionization epoch.

%
%
%A new solution with also a variable SFE in time will be presented in future works. As discussed in Section \ref{Sec_Cosmo}, it will allow us to properly treat the starburst regimes in particular phases of the Galactic evolution.
%
%
%The inclusion of the stellar life time to trace carbon \citet{gust2022}....
%
%
%
%%%%%%%

\begin{acknowledgements}
We thank the anonymous referee for helpful suggestions that improved this paper, as well as Angela Adamo for useful discussions. A.D. acknowledges support from the NASA ADAP grant 80NSSC22K0626 and GSFC award 80NSSC23K1098. S.S. acknowledges the support provided by CONAHCyT M\'{e}xico through the research grant A1-S-28458. 
Based on observations obtained with \textit{XMM-Newton}, an ESA science mission with instruments and contributions directly funded by ESA Member States and NASA, observations made by the \textit{Chandra} X-ray Observatory (CXO), and observations made with the NASA/ESA \textit{Hubble} Space Telescope obtained from the Space Telescope Science Institute. %All HST data used in this paper are available in the MAST archive at \href{https://doi.org/10.17909/rn22-jy61}{10.17909/rn22-jy61}.
This research has made use of 
%data obtained from the \textit{Chandra} Data Archive, 
software provided by the \textit{Chandra} X-ray Center in the application packages \textsc{ciao} and \textsf{Sherpa}, a collection of \textsc{isis} functions (ISISscripts) provided by ECAP/Remeis observatory and MIT, and %\footnote{\url{https://www.sternwarte.uni-erlangen.de/isis/}}
\textsf{Astropy}, a community-developed core Python package for Astronomy \citep{AstropyCollaboration2013}.
\end{acknowledgements}

%\textsc{xmm-sas} \citep{Gabriel2004}, \textsc{ciao} \citep{Fruscione2006}, \textsc{isis} \citep{Houck2000}, \textsc{xspec} \citep{Arnaud1996}, HEAsoft \citep{NHEASARC2014}, \textsf{NumPy} \citep{Harris2020}, \textsf{SciPy} \citep{Virtanen2020}, \textsf{Matplotlib} \citep{Hunter2007}, \textsf{Astropy} \citep{AstropyCollaboration2013}.

%%-----------------------------
%%   Bibliography
%%-----------------------------
%
%
%
%% The following lines are required when using BibTEX (strongly encouraged!):
%\bibliography{references}{}
%\bibliographystyle{aa}  % A&A bibliography style file (aa.bst)
%\bibliographystyle{aa_url}
%
%
%

\begin{appendix}

\end{appendix}
\end{document}